\documentclass[lettersize,journal]{IEEEtran}
\usepackage{amsmath,amsfonts}
\usepackage{algorithmic}
\usepackage{algorithm}
\usepackage{array}
\usepackage{subfig}
\usepackage{textcomp}
\usepackage{stfloats}
\usepackage{url}
\usepackage{verbatim}
\usepackage{graphicx}
\usepackage{cite}
\usepackage{bbding}
\usepackage{mathptmx}                  
\usepackage{fontawesome5}
\usepackage{adjustbox}
\usepackage{xcolor}
\usepackage{booktabs}
\usepackage{multirow}
\usepackage{tabularx}
\usepackage{makecell}
\usepackage[subtle]{savetrees}

\definecolor{clc}{HTML}{FFD541}
\definecolor{plc}{HTML}{9841FF}
\definecolor{rlc}{HTML}{FB7676}
\definecolor{alc}{HTML}{159DFF}
\definecolor{llc}{HTML}{9F9F9F}

\newcommand{\LLmark}{\textcolor{llc}{\small \faSquare}~}
\newcommand{\CLmark}{\textcolor{clc}{\noindent \small \faSquare}~}
\newcommand{\PLmark}{\textcolor{plc}{\noindent \small \faSquare}~}
\newcommand{\RLmark}{\textcolor{rlc}{\noindent \small \faSquare}~}
\newcommand{\ALmark}{\textcolor{alc}{\noindent \small \faSquare}~}

\usepackage{orcidlink}
\usepackage{caption}

\begin{document}

\title{V-RECS, a Low-Cost LLM4VIS Recommender with Explanations, Captioning, and Suggestions}


\author{Luca Podo \orcidlink{0000-0001-8780-6848}, Marco Angelini \orcidlink{0000-0001-9051-6972
}, Paola Velardi \orcidlink{0000-0003-0884-1499}
\thanks{Luca Podo and Paola Velardi are with the Computer Science Department of Sapienza University of Rome, Via Salaria 113, Rome, Italy (emails: [podo,velardi]@di.uniroma1.it}
\thanks{Marco Angelini is with Link Univesity of Rome, Via del Casale di San Pio V, 44, Rome, Italy (email m.angelini@unilink.it)}
}

\markboth{Journal of \LaTeX\ Class Files,~Vol.~14, No.~8, August~2021}%
{Shell \MakeLowercase{\textit{et al.}}: A Sample Article Using IEEEtran.cls for IEEE Journals}


\maketitle

\begin{abstract}
NL2VIS (natural language to visualization) is a promising and recent research area that involves interpreting natural language queries and translating them into visualizations that accurately represent the underlying data. As we navigate the era of big data, NL2VIS holds considerable application potential since it greatly facilitates data exploration by non-expert users. Following the increasingly widespread usage of generative AI in NL2VIS applications, in this paper we present V-RECS, the first LLM-based Visual Recommender augmented with explanations (E), captioning (C), and suggestions (S)  for further data exploration. V-RECS' visualization narratives facilitate both response verification and data exploration by non-expert users. Furthermore, our proposed solution mitigates computational, controllability, and cost issues associated with using powerful LLMs by leveraging a methodology to effectively fine-tune small models, like LLama-2-7B.
To generate insightful visualization narratives, we use Chain-of-Thoughts (CoT), a prompt engineering technique to help LLM identify and generate the logical steps to produce a correct answer. Since CoT is reported to perform poorly with small LLMs, we adopted a strategy in which a large LLM (GPT-4), acting as a Teacher, generates CoT-based instructions to fine-tune a small model, Llama-2-7B, which plays the role of a Student. 
Extensive experiments - based on a framework for the quantitative evaluation of AI-based visualizations and on a manual assessment by a group of participants - show that V-RECS achieves performance scores comparable to GPT-4 at a much lower cost.
We release V-RECS for the visualization community to assist visualization designers throughout the entire visualization generation process.
\end{abstract}

\begin{IEEEkeywords}
Visualization recommendation, LLM, Machine Learning for Visualization, Chain of Thoughts, Generative AI
\end{IEEEkeywords}

\vspace{-3mm}
\section{Introduction} \label{sec:introduction}
AI and Big Data development has reshaped how companies make decisions in recent years. A more data-driven approach has become increasingly crucial for companies to make informed decisions
\cite{tawil2023trends,saura2023digital}.
This introduces new challenges and complexity. The data analysis and transformation into actionable insights demand specific domain experts and economic resources. 
Furthermore, the supply of these specialized data scientists often falls short of meeting the needs of the industry\footnote{\url{https://shorturl.at/ePUXY}}.

Visualization Recommendation Systems (VRSs) try to fill the gap between non-expert users and data analysis \cite{podo2024agnostic} by providing automatic tools to support visual discovery from data across different sectors and applications. 
Many contributions have been developed in this field (e.g., \cite{hu2019vizml, luo2018deepeye, roth1994interactive, mackinlay1986automating}), some of which gained widespread adoption in well-known tools such as Tableau\footnote{\url{www.tableau.com} integrates a functionality named Suggest Me \cite{mackinlay2007show}}. To date, state-of-the-art VRSs do not achieve full automation, relying on built-in knowledge (such as perceptual rules) or user input to guide the system in generating visualizations.
Although fully automated solutions may lack the flexibility for detailed customization that expert users might require, they 
can also help overcome several limitations of existing approaches:\\
\indent{\textit{(i) Human visual perception}}, including color perception, pattern recognition, gestalt principles (e.g., proximity, similarity), cognitive load, and more. Capturing these processes' complexity and their interactions in a comprehensive set of rules is extremely challenging. Furthermore, rule-based approaches require manual updates to accommodate new scenarios.
Recent advancements in deep learning have enabled visual recommender systems to automatically learn perceptual rules from data/visualization pairs~\cite{podo2024agnostic}.\\
\indent{\textit{(ii) Required expertise}} for integrating user specifications in the VRSs, making it more responsive and aligned with individual user preferences. On the other hand, this interaction implies knowledgeable users.  Users without a data analysis or visualization background may find it challenging to provide the correct query, indicate the right chart type, or accurately interpret the data.
NL2Vis (Natural Language to Visualization) systems~\cite{setlur2016eviza, gao2015datatone, manning2014stanford} are being developed to simplify the interaction of non-expert users with complex data, by enabling them to generate visualizations through natural language queries (NLQ). However, reliance on these techniques poses a significant limitation, as it constrains users to adhere to specific formats (such as the need to clearly specify in the NLQ the type of chart and the variables to be plotted on the axes), thereby restricting the flexibility of expression in their queries. Furthermore, these approaches struggle to manage ambiguous texts, failing to produce a consistent and insightful visualization in many cases, as reported, for example, by Luo et al. in their work~\cite{luo2021natural}.

A further significant step forward in the automation of visual recommender systems is represented by LLM4VIS (Large Language Models for Visualization) \cite{wang2023llm4vis}. By leveraging generative AI in conjunction with visualization techniques, LLM4VIS allows it to query data sources and translate them into visualizations that accurately represent the underlying data, showing good capabilities in understanding and processing complex natural language instructions.  
It represents a promising direction in data visualization, offering advantages in terms of simplicity of user interaction, scalability, and ability to handle complex queries. However, even for these systems, many challenges remain open: 
\begin{itemize}
\item \textit{Interpretability and Trust:} Although LLM like Llama \cite{touvron2023llama} or GPT \cite{openai_gpt4} have shown a more advanced semantic understanding of free text \cite{cheng2024exploring}, they suffer from the lack of interpretability of the model's responses \cite{wang2023deciphering} and hallucinations \cite{jha2023dehallucinating}. 
\item \textit{Controllability and cost:} ``Large'' LLMs, those involving computationally intensive and expensive architectures containing billions, and recently trillions of parameters, also have other contraindications, such as the lack of controllability \cite{controllability} and the high computational cost\footnote{in \url{https://shorturl.at/puz24} it is shown that Llama 2 obtains similar performance to GPT-4 for a fraction of the cost on a number of tasks}, hindering their large-scale adoption in applications.  In particular, controllability has a notable impact on applications: if a certain model is based on the adaptation of a proprietary LLM, such as GPT, the evolution of the original LLM could significantly impact the adapted model. This means that developers of a model based on a proprietary LLM do not have full control of their system.

\item \textit{Adaptability and Verifiability:} Finally, although LLMs have proven great capabilities on tasks like text, images, or code generation, it has been observed that they are much less proficient on other tasks, such as data analysis and visualization. In this regard, Podo et al.~\cite{podo2024vi} observed how, even though GPT-3.5-turbo can already perform data visualization tasks, it still produces a high rate of errors and poses a problem of response verification, especially when the user is not sufficiently expert to evaluate the result. \textit{Evaluating the quality of a recommended visualization is a challenging task, and if users do not know what to expect, or they are not domain experts, they may perceive the model's output as accurate, even when incorrect.} 
\end{itemize}

The combination of these issues can lead to situations where the user may undertrust or overtrust the model's generated output.
To better support AI-based visualization generation while addressing its reported issues, in this paper we propose V-RECS, a LLM-based Visualization Recommender enhanced with three types of \textit{narratives}: Explanations (E), Captioning (C), and Suggestions for further data exploration (S). While E and C mitigate the problem of response verification, better supporting the trustability of the generation process for a user,  S supports non-expert users to further explore the data and fulfill data analysis tasks. The generated model,  based on a fine-tuned ``small'' LLM, Llama-2-7B, also mitigates cost and controllability issues\footnote{Llama-2-7B has seven billion parameters. Furthermore, Llama-2-7B is available free of charge for research and commercial use}. 

To achieve this, we also contribute V-RECS' training methodology. It is based on Chain of Thoughts (CoT), a strategy to instruct a large LLM  to break down its responses into reasoning steps~\cite{wei2022chain}. Since CoT has been shown not to work well with small LLM~\cite{wang2023llm4vis}, we adopt a teacher-student paradigm in which GPT-4, the teacher, is asked to enrich with visualization narratives an existing dataset of users' queries and corresponding charts, leveraging the CoT strategy. Next, the enriched dataset is used to fine-tune in a supervised manner a small open-source LLM, Llama-2-7B, to perform visualization recommendations along with producing textual narratives to explain and further explore the input data. 
V-RECS improves the generation of insightful visual representations from data while ensuring that the user has full control over the reliability of what is suggested by the Visual Recommender.\\
\indent{Summarizing}, the main contributions of this work are:
\begin{itemize}
    \item V-RECS, the first LLM fine-tuned for the visualization field that performs 
    automated visualization recommendation along with the production of insightful visualization narratives;
     \item Low-cost and controllable solution: V-RECS is based on fine-tuning the open-source Llama-2-7B, 
     presenting previously summarized advantages of low-parameter Open-AI models~\footnote{amongst the many studies, see e.g.,\url{https://tinyurl.com/56kpjd7r}}. Despite leveraging a small LLM, V-RECS achieves performances comparable to GPT-4, as shown in Section \ref{sec:evaluation}. 
    \item LLM4VIS methodology: we present an end-to-end solution for LLM4VIS, contributing a novel approach leveraging contributions from generative AI, explainable machine learning, and the visualization field.   
    \item A comprehensive comparative quantitative and qualitative evaluation experiment to study how V-RECS and GPT-4 perform the visualization recommendation task. 
\end{itemize}
We release V-RECS  and its documentation on a GitHub repository\footnote{\url{https://github.com/lucapodo/V-RECS.git}}.

The paper is organized as follows: in Section \ref{sec:LLMs} we summarize the main concepts of LLM. Section \ref{sec:related_works} is dedicated to the state-of-the-art in NL2VIS and LLM4VIS. Section \ref{sec:problem_formulation} presents a formal description of the LLM4VIS problem. Section \ref{sec:method} describes the proposed V-RECS model and the LLM-based methodology to develop it.  Section \ref{sec:evaluation} reports the results of quantitative and qualitative evaluation activities. Finally, we present a discussion and our concluding remarks in Section \ref{sec:discussion}.

\section{Background: LLM training techniques} \label{sec:LLMs}



LLMs are powerful machine learning models based on the concept of pre-training followed by fine-tuning. During pre-training, the model learns representations of language through supervised learning on extremely vast corpora of textual data. This phase is crucial for the acquisition of general linguistic knowledge. Pre-training is based on Transformer-based architectures \cite{lin2021survey}, capable of capturing long-range dependencies and contexts, essential for understanding the complexities of human language.
Following pre-training, fine-tuning tailors the LLM to specific downstream tasks by exposing it to labeled data and adjusting its parameters through supervised learning. This process adapts the model's learned representations to the specific task at hand, enhancing its performance. Recently, LLMs have been adapted to process and generate different types of unstructured data beyond text, such as images and videos, music, climate data, code, and visualizations, which is the main focus of this paper.

Another relevant concept in LLMs is Prompt engineering \cite{sahoo2024systematic}, a technique involving the design of effective instructions (prompts) or input examples to guide the behavior of language models during fine-tuning or inference. It aims to design instructions for the LLM that elicit the desired responses from the model while minimizing ambiguity and bias. Prompt engineering is a key area of research to extend the capabilities of state-of-the-art LLM. 
We can distinguish between two main prompting techniques: zero and few-shot prompting. Zero-shot \cite{kojima2022large} represents a paradigm where the model is tasked with performing a given objective solely based on a task description without explicit examples illustrating how to accomplish it. This methodology leverages the model's embedded knowledge to generate appropriate responses. 
Conversely, few-shot learning \cite{reynolds2021prompt} enhances the LLM capabilities by augmenting the task description with a small set of example inputs paired with their corresponding outputs. 

Chain of thoughts (CoT) \cite{wei2022chain} is a prompting technique to generate explanations or to perform complex reasoning tasks. It pushes the model to emulate human-like cognitive processes by telling it to break down complex problems into more manageable sub-problems, enhancing response accuracy. CoT can be classified as zero-shot \cite{kojima2022large} and few-shot \cite{wei2022chain}. The few-shot configuration necessitates a set of question-answer pair examples in addition to the new one. In this case, the prompt typically comprises multiple question/reasoning-steps/answer triplets, guiding the model through problem-solving. Unlike the few-shot variant, zero-shot CoT does not provide examples of the problem-solving process but only relies on the model's embedded knowledge.
This configuration requires a specific text formulation within the prompt ``Let's think step by step'', telling the model to split the problem into sequential sub-tasks. For instance, a query like \textit{``Problem: how much is 2+2? Let's think step by step''} drives the model to break the question and solve each part individually, culminating in a final response, e.g.: \textit{Step 1: Identify the operation to be performed: multiplication;
Step 2: Multiply the numbers: 2 × 2 = 4;
Step 3: The final result is 4.}

In our proposed framework we employ both prompting and supervised fine-tuning to tune our model. As explained in Section \ref{sec:method}, we adopt a CoT zero-shot prompting strategy to instruct GPT to generate rich visualization narratives VN when given a data table D, a NL query Q, and a visualization specification V. These narratives are used to enrich an existing dataset, which is then used to fine-tune in a supervised manner LLama. We call V-RECS the fine-tuned model. At inference time, when  D and Q are provided in input, V-RECS produces a recommendation V along with a narrative VN in zero-shot.


\section{Related work} \label{sec:related_works}
A Visualization Recommendation System (VRS) is a system designed to suggest insightful visualizations given a dataset. As outlined by Podo et al.~\cite{podo2024agnostic}, there are two classes of VRS: task-agnostic and task-aware. The task-agnostics \cite{hu2019vizml, dibia2019data2vis} require solely the dataset as input without additional guidance to generate insightful visualizations. 
Conversely, task-aware VRS requires user input or guidance, typically as user queries, to generate visualizations.

While attaining full automation in visualization recommendation pipelines would signify a remarkable milestone, it is still in the early stages, facing several challenges that have not yet been explored \cite{podo2024agnostic}.
Differently, task-aware VRSs are extensively studied in the literature. Natural Language to Visualization (NL2VIS) systems fall within this category.  In NL2VIS, the system is guided by a natural language query that tells the model what to explore in a dataset (e.g.: \textit{Which product lines generate the most revenue?}). 
The development of NL2VIS has witnessed significant advancements, primarily through classical NL techniques.
Various approaches have been proposed \cite{setlur2016eviza, srinivasan2023bolt, narechania2020nl4dv}, incorporating  NLP techniques and optional rule-based systems to recommend visualizations through an interactive visualization tool.
While these approaches provide reliable solutions, they impose input constraints by asking users to adhere to specific input templates rather than freely using natural language (for example, explicitly naming the variables to be considered in the chart, avoiding semantically ambiguous sentences, etc.). 
A departure from these syntactic approaches is explored through more advanced NLP methodologies, such as Transformers, marking the initial step toward the LLM4VIS field. NcNet\cite{luo2021natural} exemplifies these advancements by utilizing a transformer-based architecture and framing the NL2VIS problem as a translation task.
Despite being based on an advanced  architecture, due to the limited syntactic and semantic variability of the dataset used for training - compared to an LLM training dataset - the authors note that
challenges still persist in handling ambiguous and complex natural language queries.
This is a limitation, as not understanding the semantics of the query does not allow the system to understand what the user wants to explore.
In contrast, LLMs build on a much broader linguistic knowledge, demonstrating a remarkable ability to understand human language and addressing the challenge of ambiguity inherent in user expressions, which limits current approaches. \textit{Moreover, while LLMs can be fine-tuned for specific tasks, such as data visualization with narratives (as proposed in this paper), enhancing the capabilities of classic NLP models would require a system redesign.
To the best of our knowledge, none of the ''traditional" NL2Vis approaches already incorporates the ability to produce visualization narratives alongside the recommended visualization.} 

Within the field of NL2VIS, Large Language Models for Visualization (LLM4Vis) are emerging as the new frontier for translating user queries into consistent visualizations. LLM4VIS seeks to integrate LLMs into data visualization pipelines to achieve full support for visualization automation. They have shown an increasing proficiency in comprehending the underlying semantics of human instructions \cite{hadi2023large}, representing a compelling opportunity to overcome the limitations inherited by the ``classic'' approaches. While this technology is still in its early stages and requires further exploration and study, it nonetheless presents an emerging opportunity for the visualization field, as discussed by Schetinger et al.~\cite{Schetinger2024}.

We can categorize the LLM4VIS approaches into two main classes. The first class encompasses all the methods that involve the model in its standard configuration (the so-called ``plain foundation models'') and use prompt engineering techniques to improve the model's performance (Section \ref{subsec:promptunedLLM}). Conversely, the second class includes all the approaches that involve a ``fine-tuning'' phase to specialize the foundation model toward a specific set of tasks (Section \ref{subsec:finetunedLLM}). 
In what follows, we summarize the state of the art (SoA) for each category of systems. For completeness, we also dedicate a section to LLM-based systems capable of generating visualization narratives, but not visualization recommendations (Section \ref{subsec:narrativeLLM}).

\subsection{LLM4VIS prompt tuning approaches} 
\label{subsec:promptunedLLM}
A novel method for translating natural language queries into visualization is proposed by Chen et al.~\cite{chen2022nl2interface}. Their approach involves a two-step strategy: first, translating the natural language query into SQL, and second, mapping the SQL code to Vega-Lite specifications using a rule-based method \cite{chen2022pi2}. The authors leverage Codex - a pre-trained LLM developed by OpenAI for natural language to code translation\footnote{https://openai.com/blog/openai-codex} - in a few-shot prompting setup. Codex is provided with examples of natural language to SQL (NL-SQL) pairs, along with the new user query. Based on this input, the model predicts the corresponding SQL, which is then converted into a visualization.
Maddigan et al.~\cite{maddigan2023chat2vis} present a comparable study based on Codex, GPT-3, and ChatGPT to evaluate the quality of these models in generating visualizations as Python code in a zero-shot prompting setup. These papers aim to provide an initial overview of the LLM opportunities in data visualization. However, they lack a thorough quantitative and qualitative evaluation of the proposed methodology.
Li et al. \cite{li2024sheetcopilot} introduce Sheetcopilot, an LLM-based framework designed to enhance user interaction with spreadsheet tools like Excel. This framework enables users to make requests in natural language, which are then translated into spreadsheet commands using GPT-4 to achieve the desired outcome. While Sheetcopilot can handle data visualization tasks through chart generation commands, it lacks flexibility and customization in creating visualizations, as it is constrained by the predefined chart structures within the software.
A similar approach is Datacopilot introduced by Zhang et al.~\cite{zhang2023data}. Datacopilot generates automated interfaces and autonomous workflows for human-table interaction, addressing various data-related activities such as acquisition, processing, forecasting, table manipulation, and visualization. It leverages GPT to create and implement a data processing workflow with minimal input guidance. While both Sheetcopilot and Datacopilot can create visualizations as an additional feature, they are not optimized for this task. Their primary focus is on deploying interaction workflows and generating instructions or commands to fulfill user requests within software environments like spreadsheets, rather than optimizing for insightful visualizations.

Overall, while existing prompt tuning approaches offer cost-effective deployment at inference time, they come with limitations. Relying solely on prompt engineering results in a lack of control over the model's behavior in task execution. Additionally, the majority of these methods depend on proprietary models like GPT. Any updates made to these models may introduce new behaviors, potentially undermining the efficacy of the proposed methodologies. An additional concern is privacy. Relying on proprietary models makes it challenging to ensure data protection due to the inability to deploy these models privately.

\subsection{LLM4VIS fine tuning approaches.} 
\label{subsec:finetunedLLM}
A distinct approach to enhancing a Large Language Model (LLM) for visualization tasks involves fine-tuning the model. Zha et al. \cite{zha2023tablegpt} introduce TableGPT, an LLM designed to understand and operate on data tables using natural language input. Unlike other approaches, TableGPT relies on a fine-tuned model to predict a set of data analytics commands, such as question answering, data manipulation,  visualization, and report generation, based on users' input.
Moreover, Zha et al. propose a novel method to embed large data tables within the limited input length of LLMs. They employ an encoder to preprocess the table, extracting its vector representation. This embedded representation, along with the user query, is then fed into the LLM to predict the optimal command sequence.

While this approach can generate visualizations, it primarily focuses on producing tailored command sequences for performing tasks on data tables. In contrast, visualization recommender systems aim to understand the relationships between queries, data, and optimal visualization configurations, adhering to best practices in data visualization.
In line with this objective, Tian et al.~\cite{tian2023chartgpt} propose ChartGPT, a LLM-based model strictly tailored for visualization recommendation. The authors fine-tuned the open-source model FLAN-T5-XL \cite{chung2022scaling}, by splitting the data visualization generation problem into two broad tasks (i.e., data transformation and visualization recommendation). 

\subsection {LLM methods for generating visualization narratives}
\label{subsec:narrativeLLM}
 
Once a model produces a visualization or provides a response to a user query, end users face considerable difficulty in determining the accuracy of the provided output. Compared to textual responses, verifying visualizations poses a more difficult challenge. While tasks such as text summarization, content generation, or code generation allow for relatively straightforward output verification by end-users, the same cannot be said for visualizations because non-expert users may inadvertently consider misleading representations as accurate.
Consequently, the production of a ``visualization narrative'' (a description of why a recommended visualization fits a user's intentions implied by the natural language query)  emerges as crucial. The task of generating a visualization narrative delivers a textual elucidation of the underlying process behind the visualization generation process performed by the model. 
This not only enhances transparency but also reduces the potential for misinterpretation or erroneous acceptance of visualized data.

A first approach in this direction has been proposed by Sultanum et al.~ \cite{sultanum2023datatales}. The authors propose Datatales, a method to generate a visualization narrative 
when given a visualization and an annotation as input. Once the system receives the input, it is passed to GPT-3, which provides a textual narrative based on a simple prompt. Another similar method  is proposed by Ko et al. \cite{ko2023natural}. The authors present a VL2NL framework to generate a visualization narrative given a VegaLite specification. It accepts as input solely the specification and produces a visualization description and a data description as output.
Additionally, it provides a second output, representing a possible query that a generic user could use to generate the visualization. 
   
\textit{Note that both Datatales and VL2NL are tasked to generate a visualization narrative when provided with a chart or visualization specification. Therefore, they do not belong to the category of visual recommenders}. 

\subsection{Our contribution} \label{sec:contribution}
\begin{table*}[!ht]
\centering
\caption{Comparison of V-RECS and other LLM  approaches in the literature. Note that ncNet was included for completeness, as it is based on a Transformer rather than a LLM.  
}
\label{tab:related works}
\resizebox{0.99\textwidth}{!}{%

\begin{tabular}{@{}llcllccc@{}}
\toprule
\textbf{Paper}                                    & \textbf{Training strategy} & \textbf{Open Weights} & \textbf{Model} & \textbf{Model Size} & \multicolumn{1}{l}{\textbf{Visualization Narrative}} & \multicolumn{1}{l}{\textbf{Visualization recommendation}} & \multicolumn{1}{l}{\textbf{Output format}} \\ \hline
\textit{Nl2Interface \cite{chen2022nl2interface}} & Prompting                  & $\times$              & Codex          & 12B                 & $\times$                                             & \Checkmark                                                & code                                      \\
\textit{Chat2Vis \cite{maddigan2023chat2vis}}     & Prompting                  & \textit{$\times$}     & GPT-3          & 175B                & $\times$                                             & \Checkmark                                                & code                                    \\
\textit{ChartGPT \cite{tian2023chartgpt}}         & Fine-tuning                & \Checkmark            & Flan-t5-xl     & 3B                  & $\times$                                             & \Checkmark                                                & grammar                                    \\
ncNet \cite{luo2021natural}                       & Training                   & \Checkmark            & Transformer         & n.a.                   & $\times$                                             & \Checkmark                                                & grammar                                    \\
\textit{Data-copilot \cite{zhang2023data}}        & Prompting                  & $\times$              & GPT-4          & 1T                  & E                                                    & \Checkmark                                                & code                                      \\
\textit{TableGPT \cite{zha2023tablegpt}}          & Fine-tuning                & \Checkmark            & Phoneix        & 7B                  & C                                                    & \Checkmark                                                & commands                                      \\
\textit{SheetCopilot \cite{li2024sheetcopilot}}   & Prompting                  & $\times$              & GPT-4          & 1T                  &    $\times$                                                 & \Checkmark                                                & commands                                      \\
\textit{VL2NL \cite{ko2023natural}}               & Prompting                  & $\times$              & GPT-4          & 1T                  & ECS                                                  & $\times$                                                  &   n.a.                                  \\
\textit{Datatales \cite{sultanum2023datatales}}   & Prompting                  & $\times$              & GPT-3          & 175B                & EC                                                   & $\times$                                                  & n.a.                                    \\ \midrule
\textit{\textbf{V-RECS}}                          & \textbf{Fine-tuning}       & \Checkmark   & \textbf{Llama} & \textbf{7B}         & \textbf{ECS}                                 & \textbf{\Checkmark}                                       & \textbf{grammar}                           \\ \bottomrule
\end{tabular}
}
\end{table*}

Table \ref{tab:related works} summarizes the main features of the previously surveyed LLM-based systems. For completeness, we also include NcNet since, although not based on an LLM, it leverages a Transformer architecture. Also note that other traditional NL2Vis approaches are not included in the comparison since, unlike LLMs, they cannot be adapted to generate visualization narratives; rather, they should have been designed with these features from scratch.

To compare the systems with each other and with the solution proposed in this paper, we consider seven dimensions: 
\begin{enumerate}
\item \textit{Training strategy:} specifies whether the model uses training from scratch, fine-tuning (pre-trained model aligned to specific tasks), or prompting (tuned at inference time with a prompt).
\item \textit{Model}, identifies the LLM model used in the proposed strategy (e.g., GPT, Llama).
\item \textit{Open Wieghts :} Indicates whether the model weights are publicly available (\Checkmark) or proprietary ($\times$). Public weights enable private deployment and allow for new fine-tuning strategies, providing more control over the model. In contrast, proprietary models require API interaction and data management on external servers, which can be a limitation for sensitive data and restrict users from using the provided methods to enhance/refine the model.
\item \textit{Model size:} Along with open weights, the model size is crucial for private deployment. Smaller models like Llama 7B can be easily deployed on commercial computers, while larger models require more computational resources, posing a limitation for private deployment.
\item \textit{Visualization recommendation :} this dimensions tells if the system performs (\Checkmark) or does not perform ($\times$) the task.
\item \textit{Visualization narrative :} this dimension specifies which types of narratives are generated (E, C, S), while $\times$ means that no narratives are produced.
\item \textit{Output Format:} this column compares visualization methods based on how they generate visual outputs. Some methods provide visualization as code or grammar, allowing for highly customizable and flexible visual representation. On the other hand, tools that generate predefined commands for visualizations offer a more guided -but less flexible- approach, where users utilize pre-packaged commands to create visual outputs. Note that, since VL2NL and Datatales do not generate visualizations, this classification is not applicable (n.a.).

\end{enumerate}

Hereafter, we propose V-RECS, a model capable not only of generating a coherent visualization, as proposed by Tian et al.~\cite{tian2023chartgpt}, but also of explaining the generated output by integrating narrative visualization capabilities, as suggested by Ko et al. ~\cite{ko2023natural}. Unlike the few systems that produce visualization narratives in the form of captions, V-RECS generates a very rich narrative, composed of a detailed explanation of the reasoning process that led to the production of the visual recommendation, a caption to describe the recommended chart, and suggestions for further exploration of the data.
Furthermore, V-RECS leverages a methodology to adapt smaller models to perform as the larger ones, as demonstrated in Section \ref{sec:evaluation}. Remarkably, V-RECS' fine-tuning strategy exploits a ``small'' and open-source LLM model, thus mitigating cost, privacy, and controllability issues mentioned in Section \ref{sec:introduction}.

\section{Problem formulation} \label{sec:problem_formulation}

We can formulate the goal of LLM4VIS systems as maximizing the probability of suggesting the best visualization to a user, given a data table $T$ and a user query $Q$:
\begin{equation}
    \centering
    f_{i}(Q_{i}, \mathfrak{T})\rightarrow argmax_{V_{i}}P(V_{i}|\mathfrak{T},Q_{i})
    \quad where \quad V_{i}:\{v_{i,j} |0 < j\leq n \}
\end{equation}

where $Q_{i}$ is the user query at time $i$, $\mathfrak{T}$ is the table representation passed to the model and $V_{i}$ is the output visualizations set (a single visualization or a dashboard). 
Note that, commonly,  $V$ is not directly a chart, but rather, the specification of a chart in a visual representation language, typically, a JSON-based language like Vega-Lite and the simpler VegaZero \cite{luo2021natural}. 

\begin{figure*}[!t]
    \centering
    \includegraphics[width=\textwidth]{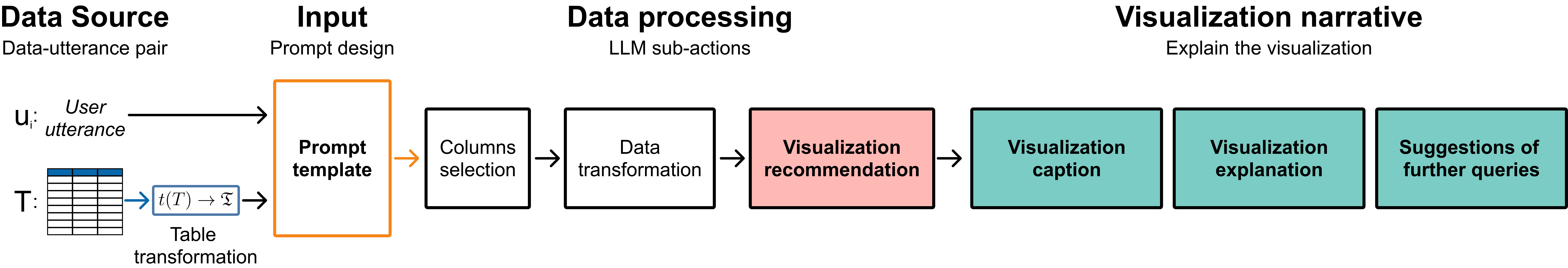}
    \caption{V-RECS model task workflow. The three green boxes represent V-RECS' specific tasks, while the others are common to all LLM4VIS models.}
    \label{fig:task}
\end{figure*}

The previous is a general formulation that also applies to NL2VIS systems. LLM4VIS systems introduce a series of intermediate tasks, illustrated in Figure \ref{fig:task}.
First, it is crucial to note that the model does not directly receive the original data table $T$ as input, but rather $\mathfrak{T}$, which results from a \textit{table transformation} operation. 
Given the LLM's limited input size, depending on the number of input tokens a model can accept, it becomes essential to pre-process the original table into a format suitable for model input. 
To face this problem, for example, Zha et al.~\cite{zha2023tablegpt} transform the table using an encoder; differently, Tian et al. ~\cite{tian2023chartgpt} propose to use only the table description with some rows to provide the model a better context of the data. Following the table transformation, the input is shaped by a \textit{prompt function} $P(u_{i}, \mathfrak{T})$ (see Section \ref{sec:LLMs}) that maps the query and table representation to a designed prompt template. 
After that, the LLM should perform the following tasks to produce $V$ (as shown in Figure \ref{fig:task}, left):
\begin{itemize}
    \item \textit{columns selection}. Once the model receives the prompt as input, the model has to determine the most meaningful and prominent columns from the data table based on the user query.
    \item \textit{data transformation}. After the model has chosen the columns based on the user query and the column's data type, the most appropriate data transformation functions should be selected. For instance, the model should determine whether to aggregate (e.g., min, average, max) different data columns or values.
    \item \textit{visualization recommendation}. Based on the output of the previous steps, the model must choose the best visualization elements (axes, marks, presentation, etc.) to map the data and generate the visualization as code (e.g., matplotlib python code to produce a bar chart), image (e.g., bar chart as pixel representation), or grammar (e.g., a visualization expressed using VegaLite specifications).
    \end{itemize}

    In addition to the above tasks, in this paper we propose that the LLM model generates a \textit{visualization narrative} 
 VN, which we split into three tasks  (see the green boxes in Figure \ref{fig:task}, and the right-bottom side of Figure 1 for an example): 
    \begin{itemize}
      \item \textit{Explanation (E)}. First, the model should explain why specific columns have been selected from the available dataset, and which transformation functions (e.g., average) have been applied. This transparency is essential for users to understand the decisions behind model choices, allowing them to verify the accuracy of the generated visualization and its adherence to users' information needs.
    \item \textit{Captioning  (C)}. Subsequently, the model should generate descriptive captions for the produced visualizations. These captions elucidate the visual representation's content and meaning, helping the user understand the data shown.
  
    \item \textit{Suggestion (S)}. Finally, the model should offer suggestions for additional data queries, guiding users on where to delve deeper into the dataset for further insights. 
\end{itemize}
\noindent 
We name V-RECS the specialized LLM4VIS model able to generate visualization recommendations along with the above tasks, referred to as  \textit{generation of visualization narratives}.
Accordingly, the formal specification of the V-RECS model task is as follows:
\begin{equation}
    \centering
    f_{i}(Q_{i}, \mathfrak{T})\rightarrow argmax_{V_{i}}P(V_{i}, VN_{i}|\mathfrak{T},Q_{i})
    \quad where \quad V_{i}:\{v_{i,j} |0 < j\leq n \}
\end{equation}
where $VN_{i}$ denotes the rich narrative accompanying the generated visualization.




\section{V-RECS Model} \label{sec:method}
\begin{figure*}[!ht]
    \centering
    \includegraphics[width=.85\textwidth]{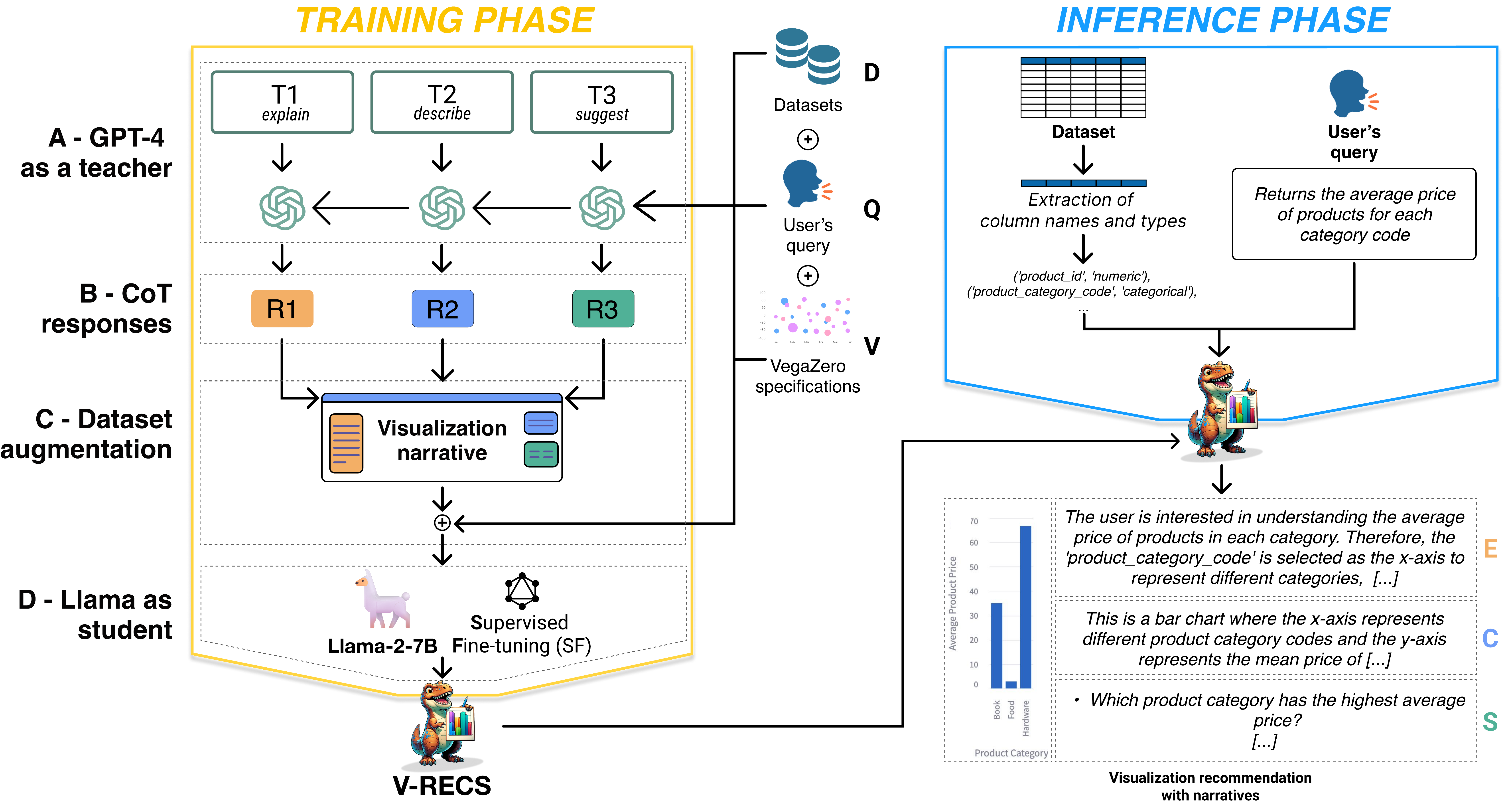} 
    \captionof{figure}{During the \textit{training phase} (left), a large LLM, GPT-4, acting as a Teacher (\textbf{A)}, receives as input triples (D,Q,V) extracted from an available dataset, where D is tabular data, Q is a natural language query, and V is the VegaZero specification for a chart matching the user's query Q. The model is prompted to perform three Chain of Thought (CoT) tasks: T1, to explain the reasoning steps that justify the answer V given the query Q; T2, to generate a caption that describes the chart, and T3, to suggest additional useful queries. The responses of the model are combined to create a visualization narrative (VN) used to augment the initial dataset (\textbf{C)}. Next,  a small model, Llama-2-7B (the Student), is fine-tuned (\textbf{D)} with the enriched dataset of quadruples (D,Q,V,VN). At \textit{inference time} (right),  the resulting specialized model, named V-RECS, receives as input a pair (D,Q) and generates, along with a visualization recommendation V, a visualization narrative VN made of an explanation E, a caption C, and a suggestion  S of additional queries.    
    Adopting a teacher-student metaphor, the esteemed professor GPT-4, leveraging its trillion-parameters knowledge of data visualization and natural language understanding, generates examples of visualization narratives for the young student Llama-2-7B. Llama uses this supplementary material to diligently learn the task of producing similar narratives alongside the recommended visualization, and finally gets its doctorate in Visual Recommendation with Narratives, becoming Dr. V-RECS. During the exercise of its profession, will Dr. V-RECS be able to match Professor GPT-4's abilities? Certainly, its professional fees are much cheaper.
    }
    \label{fig:pipeline}
\end{figure*}

In this section, we describe the methodology to implement V-RECS. 
The proposed methodology is based on fine-tuning a small model in a supervised manner, using an existing dataset enriched with visualization narratives generated by a more powerful model.  
The methodology is depicted in Figure \ref{fig:pipeline}. First (box A of Figure  \ref{fig:pipeline}, left), GPT-4, acting as a Teacher,  is instructed to generate visualization narratives leveraging the
Chain-of-Toughts zero-shot prompting technique proposed by Ko et al. \cite{ko2023natural} (see Section \ref{sec:LLMs}). The Teacher receives input from an existing dataset composed of triples (D,Q,V) where D is tabular data, Q a user's query in natural language, and V a visualization specification matching the user's needs expressed by Q. Based on the CoT approach, the Teacher is prompted with three CoT tasks, each with a different goal. The first task, T1, is to explain why the visualization V is a correct answer, given the user's query Q, while the goal of T2 involves generating a caption to describe the chart. Finally, the goal of T3 is to suggest potential areas of interest for further data exploration to the user.  
The Teacher, based on the three prompts, generates the responses for the three tasks, $R_1$, $R_2$, and $R_3$ (box B). Next, the responses are combined, along with the original triple  (D,Q,V), to create training instances for the subsequent fine-tuning phase (box C).
During fine-tuning (box D), the training instances - which are now composed of sextuples (D,Q,V,E,C,S) -
are used 
to teach a smaller LLM model (the Student LLama-2-7B) to generate visual recommendations along with visualization narratives. At inference time (right-hand side of Figure \ref{fig:pipeline}), the fine-tuned model, named V-RECS (Visualization Recommender with Explanations, Captions, and Suggestions), receives as input a (D,Q) pair and recommends a visualization V along with a  narrative VN composed by an explanation (E), a caption (C), and suggestions for further explorations (S). 

To instruct the models we use the zero-shot prompting strategy 
for the Teacher and fully supervised training to fine-tune the Student. At inference time instead, V-RECS works in zero-shot.
In what follows we explain the methodology in more detail.

\subsection {Model training methodology}
As described in Section \ref{sec:problem_formulation}, the goal of V-RECS is to recommend visualizations, along with rich narratives to explain and extend recommended visualizations. Furthermore, we also aim to build our solution by leveraging a small and open-source LLM model, the advantages of which we have already described in Sections \ref{sec:introduction} and \ref{sec:related_works}.
Smaller models specialized in a specific task, with comparable performance to larger ones, represent a terrific opportunity to reach the same results at a minor cost while leveraging an open source model. 

To enrich the explainability capabilities of our model and instruct it to generate rich visualization narratives, we leverage Chain-of-Thoughts, a reasoning approach that mimics human problem-solving by breaking down complex tasks into simpler, sequential stages (see Section \ref{sec:LLMs}). 
CoT has demonstrated substantial advantages, particularly in inferring new tasks with minimal effort, without necessitating structural changes to the underlying model architecture \cite{wei2022chain}. However, its application to models with fewer than 100 billion parameters has been associated with logical inconsistencies during the reasoning process \cite{wei2022chain}.
To address these discrepancies, fine-tuning smaller models emerges as a viable solution to enhance the proficiency of small models in solving complex tasks, such as data visualization. Nonetheless, this approach presents its limitations. Fine-tuning smaller models on complex tasks using question-answer pairs alone, devoid of reasoning steps, often produces incorrect responses. 

To overcome these limitations and augment the capabilities of smaller models in data visualization tasks, we propose leveraging a novel approach known as \textit{Fine-tuning based on CoT} \cite{ko2023natural}. This method combines both CoT and fine-tuning by enriching question-answer pairs without reasoning steps with the sub-tasks needed to solve the problem posed, improving the reliability of a smaller model in producing correct outputs \cite{bursztyn2022learning}. The process involves exploiting the embedded knowledge of a larger
``Teacher'' model, such as GPT-4, in a zero-shot  CoT configuration, to provide responses enriched with reasoning steps. Subsequently, a smaller ``Student'' model is fine-tuned based on these responses, effectively leveraging the Teacher's knowledge.

While fine-tuning based on CoT has been used to augment datasets for solving scientific problems \cite{lu2022learn}, its application to visualization tasks remains unexplored. 
Building upon the methodology by Ko et al. \cite{ko2023natural}, we have devised a custom CoT strategy for the V-RECS problem domain, involving three ``Teachers,'' each tasked with addressing specific challenges related to the V-RECS tasks, including visualization recommendation and captioning, explanation of the process of column selection and filtering in D, and suggestion of additional questions for data exploration. 
The Teachers' tasks are delineated in Table \ref{table:teachers}, with variations based on the input and intended goal.

Finally, the responses generated by the three Teachers, along with the original query Q 
and the expected visualization V are combined to fine-tune the Student, as detailed below. 
\begin{table}[!h]
\centering
\caption{Description of the Teacher's  tasks}
\label{table:teachers}
\resizebox{\columnwidth}{!}{%
\begin{tabular}{llll}
\hline
Task & Input    & Goal     & Description                                                                                   \\ \hline
\textbf{T1} &
  \begin{tabular}[c]{@{}l@{}}Query, \\ VegaZero, \\ Dataset\end{tabular} &
  \textit{Explain} &
  \begin{tabular}[c]{@{}l@{}}Explain to the user why the features used in the \\ VegaZero specification V are the most appropriate, given all \\ the dataset features and the user's query\end{tabular} \\ \hline
\textbf{T2}   & VegaZero & \textit{Caption} & Generate a caption to describe the content of the visualization                             \\ \hline
\textbf{T3}   & VegaZero & \textit{Suggest} & \begin{tabular}[c]{@{}l@{}}Discover the most insightful questions to \\ suggest to the user for further exploring the data\end{tabular} \\ \hline
\end{tabular}%
}
\end{table}


\subsection{Implementation details}
\label{subsec:implementation}
In this section we provide the relevant implementation details to replicate our methodology. Additional details can be found in the annexed Supplemental Material and in the V-RECS GitHub \footnote{\url{https://github.com/lucapodo/V-RECS.git}}.

\noindent
\textbf{Visualization representation}
A challenging aspect concerns the choice of a visualization description method. A visualization can be represented as a declarative specification, a bitmap image, or through its code. The first type is based on visualization grammars, like the popular Vega-Lite. The second type considers the visualization as an image, and the third type represents the visualization as a set of instructions, for example, using Matplotlib in Python. Pixel-based visualizations would add limitations and complexity to our approach, as LLMs are primarily trained using textual data, while limiting interactivity and deployability.
Similarly, employing a code-based visualization method (such as some of the systems surveyed in Table \ref{tab:related works}) would hinder the creation of a pure LLM approach, requiring a specific code environment to render and generate visualizations. Differently, a grammar-based visualization can be considered agnostic to the used rendering environment, making it more general than code, while still formatted as structured text. For this reason, we chose grammar-based visualizations as the most suitable for our scenario. 
Specifically, we have opted for VegaZero, proposed in \cite{luo2021natural}. 

\noindent
\textbf{Selection of the training dataset}
Various datasets have been proposed in the literature for Natural Language to Visualization (NL2VIS) tasks, with NvBench \cite{nvBench_SIGMOD21} emerging as the most prominent and widely adopted. The natural language queries in the NvBench dataset were generated by starting with existing NL2SQL benchmarks, modifying SQL queries into visualization queries, filtering out low-quality visualizations, and then manually verifying the quality of the synthesized (NL, VIS) pairs. Despite its widespread adoption, the dataset exhibits limitations such as a lack of complexity in the visualization types, overly explicit natural language queries (except for some instances pre-classified as ``hard''), and the absence of support for visualization narratives or reasoning steps. 
Additionally, due to the generation process described above, some of the NL queries in the dataset may not appear natural\footnote{as shown by some of the examples used throughout this paper, which are all taken from the NvBench dataset}. However, this apparent problem replicates a well-studied \cite{ungrammatical} real-world challenge, where occasional and/or non-expert users often employ informal, ungrammatical, and sketchy language. Recall that one of the motivations for leveraging natural language understanding features in visual recommenders is to target these types of users. Thanks to their advanced language processing capabilities, LLM4VIS may infer a user's information need even when expressed in complex, incomplete, or sketchy form.   

While other datasets like NLVCorpus \cite{srinivasan2021collecting} or Quda \cite{fu2020quda} provide alternatives, they also come with their own constraints. In particular, NLVCorpus includes only 893 visualization-query pairs, and Quda does not provide the chart associated with the queries.
In the absence of a dataset capable of supporting our training requirements (i.e., sufficiently large samples of D,Q,V triples), we opted to choose NVBench as the best fit for our purpose.
We chose the NvBench version provided by NcNet, as it aligns closely with our training task requirements, i.e.,  providing the queries, visualizations, and data sources. Specifically, this version excludes visualizations generated from multiple data tables and adds VegaZero representations tailored to each query. Other datasets, such as VisText \cite{2023-vistext} or VIS30k\cite{Chen2021}, do not include data-query-visualization (D,Q,V)  triples and focus solely on caption generation.

\begin{figure}[!ht]
    \centering
    \includegraphics[width=0.9\linewidth]{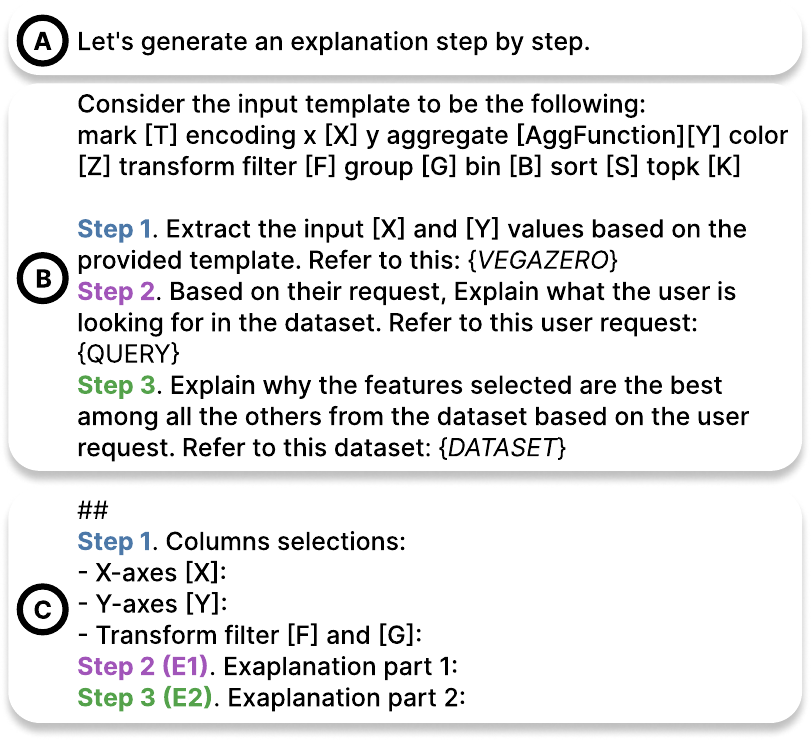}
    \caption{Structure of the T1 prompt. The related task is described in Table \ref{table:teachers}. It has three parts: \textbf{A} encourages the LLM to reason ``step-by-step'' as per the Chain of Thoughts method, \textbf{B} includes the description of sub-tasks (steps) to guide the model, and \textbf{C} is the response template. Note, in box B,  that Step 2 and Step 3 direct the system to split the explanation into two parts. Part 1 (E1) summarizes the user's information needs, to make sure they have been correctly interpreted; part 2 (E2)  explains why, given D and Q, certain features X and Y and transformation functions have been selected.}
    \label{fig:t1}
\end{figure}

\noindent
\textbf{Teacher and Student prompting}
GPT-4 has been selected as the Teacher and prompted with three distinct tasks (T1, T2, and T3), described at a high-level in Table \ref{table:teachers}, based on the input and task delivered. Furthermore, leveraging the work of Ko et al.~\cite{ko2023natural}, we provide each prompt with in-context reference to the VegaZero template to help the model generate consistent visualization.
Each task is tailored for Zero-shot CoT supplemented with embedded sub-tasks, denoted here as ``steps''. These steps act as ``guiding principles'', offering the Teacher subtle cues on approaching and unraveling the task.  The steps vary for each task.
Figure \ref{fig:t1} shows the detailed description of the prompt template for T1 (Explain), composed of three blocks and related steps. Note that the explanation is split into two parts, described by Steps 2 and 3 of Figure \ref{fig:t1} (box B). The other prompt templates, along with a more detailed description,  can be inspected in the Supplemental Material.

\begin{figure*}[!ht]
    \centering
    \includegraphics[width=0.9\textwidth]{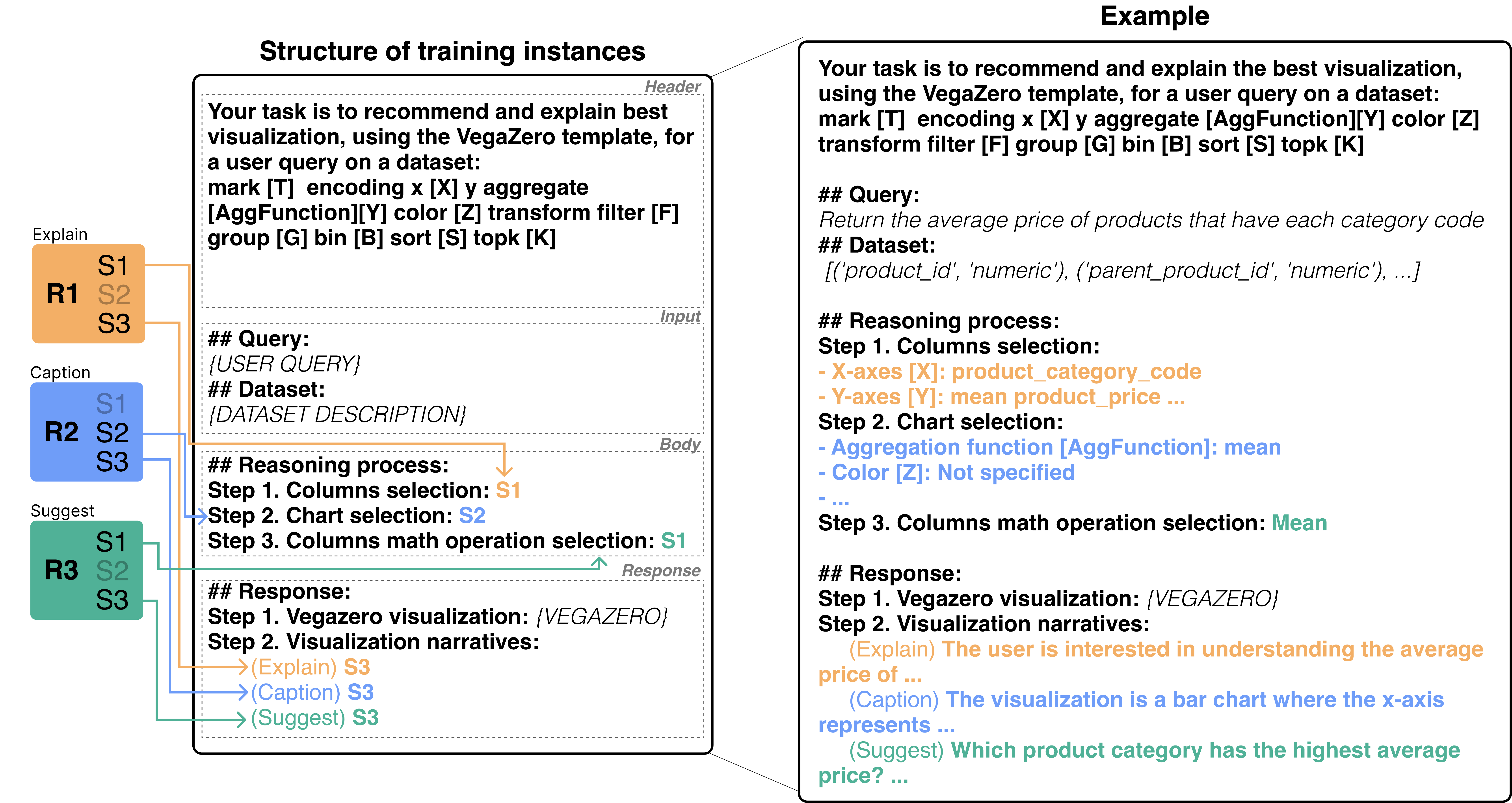}
    \caption{Training template during fine-tuning. The leftmost box shows how the Teacher's responses R1, R2, and R3   contribute to filling the different parts of a Student's training instance. The rightmost box is a (partly) filled template for a specific training instance.}
    \label{fig:template}
\end{figure*}

\begin{figure}[!ht]
   \centering
   \includegraphics[width=\linewidth]{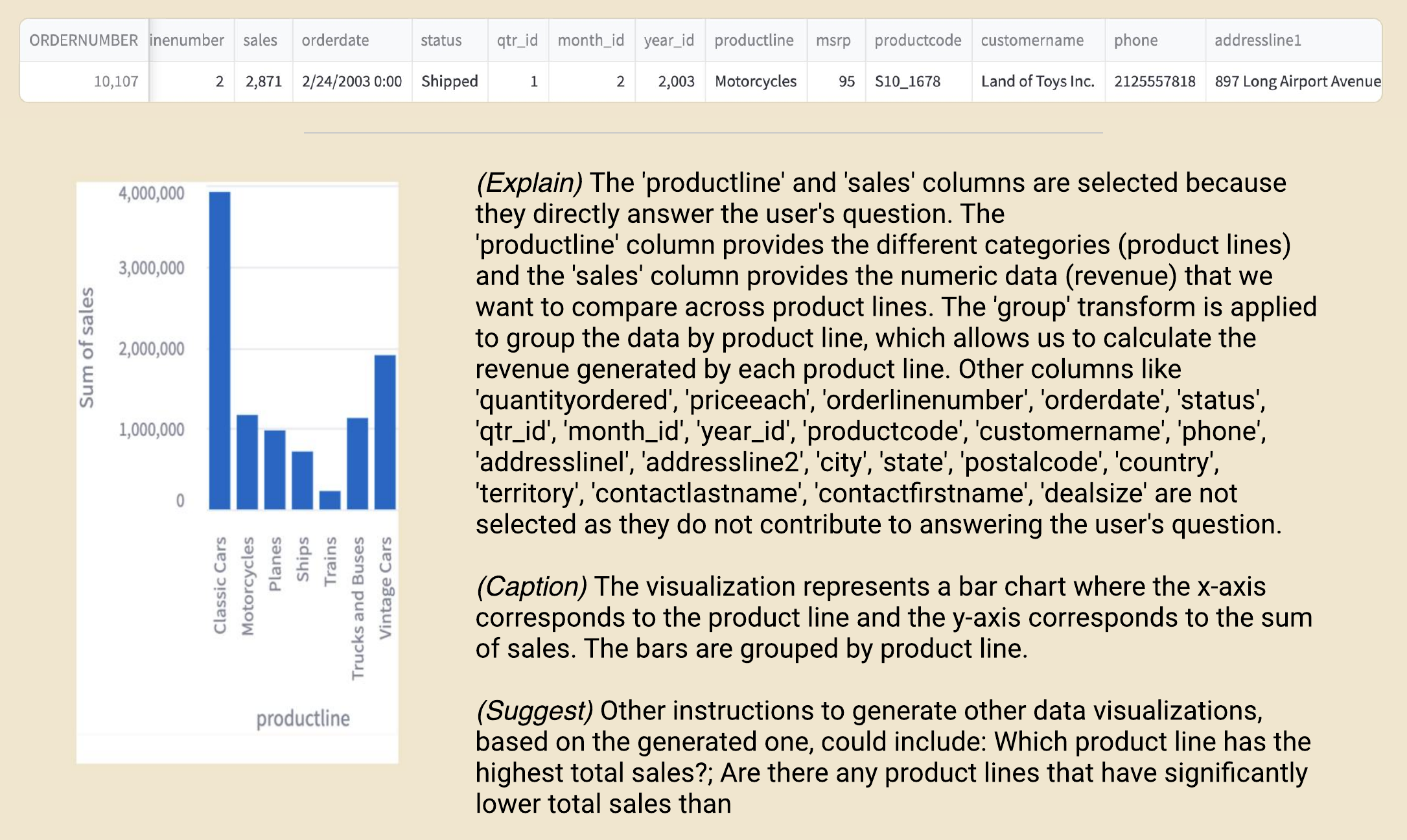}
   \caption{Example of V-RECS response at inference time for the query \textit{\textbf{Which product lines generate the most revenue?}} }
   \label{fig:response}
\end{figure}


To instruct the Teacher, we used a dataset comprising 2938 NvBench triples, each consisting of a dataset D, a user's query Q, and a VegaZero specification V, following the same distribution of query complexity of the original dataset\footnote{the complexity level is provided by the creators of the dataset. }.
Based on the three prompts, the Teacher generates, for each input triple (D, Q, V) of the dataset, the related responses $R_1$, $R_2$, and $R_3$. Next, the three responses $R_i$ along with the corresponding D, Q, and V are combined into a \textit{training instance template} used to instruct the Student, consisting of the header, the input, the body, and the response, as follows (see Figure \ref{fig:template}): 
\begin{enumerate}
\item\textbf{The header} includes the explicit task the model is supposed to pursue. It drives the model to understand what it has to achieve. Along with the task, this part also includes the VegaZero template structure.
\item\textbf{The input} includes the user query and the dataset description. The dataset is embedded using only the data feature names and related types. 
\item\textbf{The body} consists of the intermediate reasoning steps the model should follow to select proper visualization properties, such as the aggregation function. This part is crucial to provide additional information to the end-users so they can understand the model's design choices and generate the final response. Adding this part enables the explainability of the generated response. 
\item\textbf{The response} is the last part of the prompt, where we provide the structure of the answer that the model must present to the end-user. It includes the VegaZero specification V to generate a chart (the recommended visualization), and the E,C,S triple that makes up the visualization narrative.
\end{enumerate}
The training instance template, along with a (partially) filled training example, is depicted in Figure \ref{fig:template}. The left-hand side of Figure \ref{fig:template} shows how the different components of the Teachers' responses contribute to filling the template \footnote{More details on how the Teacher's responses are combined to fill the template are provided in the Supplemental Material.}. Further note in Figure \ref{fig:template}, left, the partly ungrammatical structure of the query. The query shown is included in the NvBench dataset, and, as previously remarked, mimics the tendency of non-expert users to employ a sketchy language.

\noindent
\textbf{Fine-tuning }
As a result of the Teacher CoT phase, the initial 2938 triples are enriched with the visualization narratives generated by GPT, and packed into training instances, shaped as in Figure \ref{fig:template}. The last step is fine-tuning the Student with the generated training instances.  

As a Student, we have used Llama-2-7B \cite{touvron2023llama}, the smallest from the Llama2 family. 
Drawing from the QLora technique  \cite{dettmers2024qlora}, our training approach went beyond attention modules \cite{vaswani2017attention} to incorporate linear components. QLora has shown very promising results during training. With reference to \cite{dettmers2024qlora}, when configuring Lora, we opted for an $r$ value of 64 and an $\alpha$ value of 128. During fine-tuning, we employed a batch size of 4, a learning rate of $1e^{-4}$, and utilized the AdamW optimizer.

The resulting fine-tuned model is V-RECS. The V-RECS model was deployed on AWS Sagemaker with an A10G Nvidia GPU, fed using only the user's query and the dataset. Differently, GPT-4 operated in a zero-shot CoT, with prompt mirroring used during V-RECS training. 

Figure \ref{fig:response} shows an example of V-RECS response at \textit{inference time}. As shown, during inference, V-RECS is given a dataset D and query Q and returns a visualization recommendation along with a narrative composed of the three elements E, C, and S.


\section{Evaluation} 
\label{sec:evaluation}


This section presents a comprehensive evaluation of V-RECS, encompassing quantitative and qualitative assessments on evaluating V,E,C and S. The evaluation aims to ascertain to what extent V-RECS has achieved the proposed objectives, i.e. generating insightful visualization recommendations supported by rich narratives, while relying on a non-proprietary and low-cost LLM. In both quantitative and qualitative evaluation, we compare the results generated by V-RECS with the ones generated by GPT-4. While a baseline evaluation was initially planned with Llama-2-7B (the un-tuned model), extensive experimentation with prompt templates and incremental prompt construction yielded unsatisfactory results, leading to its exclusion from further evaluation.
We emphasize that the only possible comparison is with GPT-4, the most powerful LLM to date since other LLM4Vis (and NL2Vis) systems do not produce visualization narratives, nor can they be easily adapted to do so. \textit{The ability to generate rich visualization narratives represents the main contribution of V-RECS compared to the state of the art, which would make a comparison with models without this functionality inappropriate.}

The evaluation was carried out using the EvaLLM framework, which we introduced in \cite{podo2024vi}. Similar to the popular ISO-OSI stack, EvaLLM defines a five-layer evaluation framework where each layer's output serves as the input for the layer above it, with each layer focusing on evaluating a different property of the generated visualization. 
The layers are: \CLmark The code layer serves as the foundational level, evaluating fundamental structural properties. \PLmark The representation layer addresses core visualization properties based on established representation rules. \RLmark The presentation layer assesses the perceptual quality of data presentation. 
\ALmark The informativeness layer gauges the intrinsic quality of visualizations in terms of insights and adherence to best practices in visualization literacy.  \LLmark Lastly, the LLM layer evaluates the computational and strategic costs of generating specific visualizations. Layers are further characterized into \textit{levels}, each evaluating distinct properties.
We refer the interested reader to \cite{podo2024vi} for additional details. 

To evaluate V-RECS, we focused only on four of the five layers originally proposed in EvaLLM, developing our stack implementation. Additionally, for the informativeness layer, we introduced a new level, ``Narratives quality''. 
We implemented the selected levels as described in Table \ref{tab:evallm}. Note that the first four levels in  Table \ref{tab:evallm} can be evaluated quantitatively, while both Informativeness and Effort are assessed with a qualitative evaluation. 
\begin{table}[!t]
\caption{Description of the implemented evaluation levels based on the EvaLLM framework. The layers to which every level belongs are indicated by the colored boxes. ''Ground-truth" refers to a subset of  1800 (D,Q,V) triples, generated as discussed in Section \ref{sec:method}, not used to train V-RECS.  }
\label{tab:evallm}
\fontsize{8pt}{9pt}\selectfont
\begin{tabularx}{\linewidth}{l l X}
\toprule
Levels & Metric & Implementation \\
\midrule
  \begin{tabular}[c]{@{}l@{}} \CLmark Syntax \\ correctness\end{tabular} 
& \begin{tabular}[c]{@{}l@{}} Grammar \\ accuracy\end{tabular} 
& It employs the Altair Python library to validate GPT responses and a parser developed by VegaZero authors for V-RECS. \\
\midrule 

    \begin{tabular}[c]{@{}l@{}} \RLmark Data \\ mapping\end{tabular}  
    & \begin{tabular}[c]{@{}l@{}} Data mapping \\ accuracy\end{tabular} 
    & It measures the model's ability to select appropriate columns from dataset D based on user queries Q. The metric is computed by comparing the columns selected by each model against the ground-truth columns. \\
    \midrule
    \begin{tabular}[c]{@{}l@{}} \RLmark Mark \\ correctness\end{tabular}  
    & \begin{tabular}[c]{@{}l@{}} Mark \\ accuracy\end{tabular}  
    & This level computes the metric by comparing the mark selected by each model against the ground-truth. \\
    \midrule
    \begin{tabular}[c]{@{}l@{}} \RLmark Axes \\ quality\end{tabular}  
    & \begin{tabular}[c]{@{}l@{}} Axes \\ accuracy\end{tabular}  
    & This level was evaluated based on accuracy, comparing the predicted axis mapping against the ground-truth. \\
\midrule
    \begin{tabular}[c]{@{}l@{}} \ALmark Narrative \\ quality\end{tabular} 
    & \begin{tabular}[c]{@{}l@{}} Human \\ evaluation\end{tabular} 
    & This level involves a human-based qualitative evaluation in the form of a comparative study based on a custom web app  (see Section \ref{quantitative}). \\
\midrule
\LLmark Effort 
& \begin{tabular}[c]{@{}l@{}} Computational \\ cost\end{tabular} 
& Evaluates the complexity of the strategy used to obtain the desired results, ranging from zero-shot prompting for GPT (low cost) to fine-tuning for V-RECS (high cost). \\
\bottomrule
\end{tabularx}
\end{table}
\begin{figure*}[!ht]
    \centering
    \includegraphics[width=0.95\linewidth]{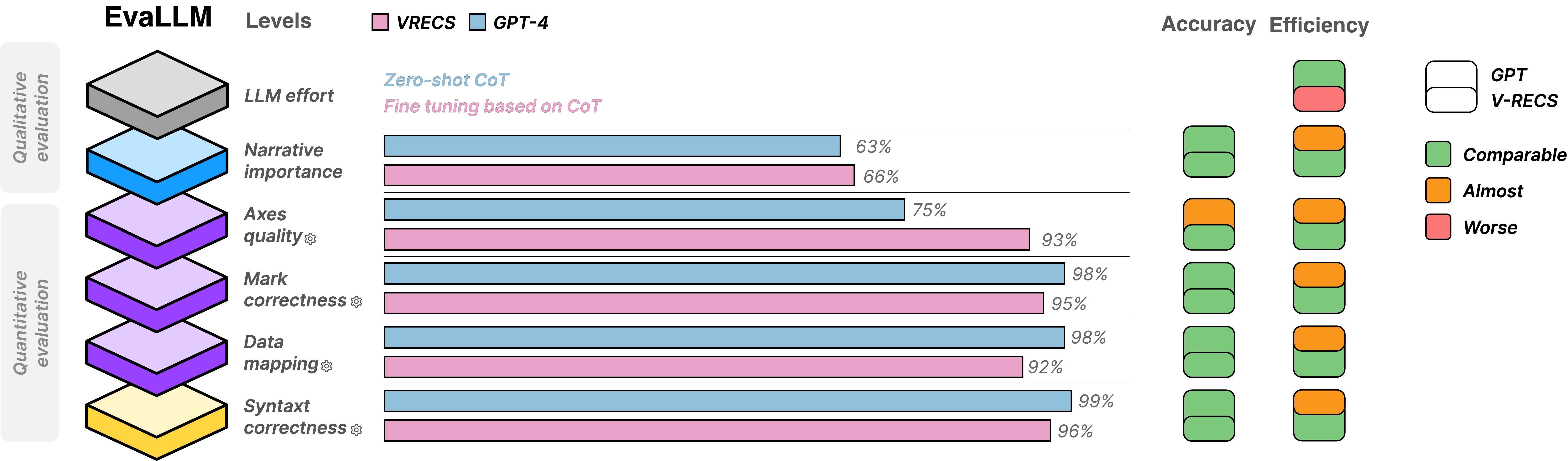}
    \caption{Results of the EvaLLM evaluation experiment}
    \label{fig:evallm}
\end{figure*}

\subsection{Quantitative evaluation}
\label{quantitative}
The quantitative evaluation involves Syntax correctness, Data mapping, Mark correctness, and Axes quality (see Table \ref{tab:evallm}).
This experiment aimed to prompt both GPT-4 and V-RECS models with a user's query and datasets, subsequently evaluating the responses generated by each model. These responses encompassed the recommended visualizations V and the accompanying narrative  (E,C,S), allowing for a comprehensive performance assessment. 
For V-RECS the prompt solely consisted of the user query, dataset, and task description.
Conversely, the prompt for GPT-4 was more elaborate, including the task, zero-shot CoT formulation, a set of sub-tasks guiding its response generation, and the desired response format\footnote{i.e., we use a template format similar to those used for the three teachers, see Section \ref{subsec:implementation}}. 

For the evaluation, we used  1800 samples drawn from the original ncNet dataset \cite{luo2021natural}, not used in the training phase,  augmented with the generated visualization narratives, as described in Section \ref{sec:method}.
As shown in Figure \ref{fig:evallm}, the two models exhibited comparable accuracy across three quantitative levels (Syntax, Data, and Mark), with GPT-4 showcasing a slight advantage from +3\% to +4\% accuracy. 
However, a notable discrepancy emerged in assessing Axes quality, where GPT-4 appeared to significantly underperform compared to V-RECS (75\% against 93\% accuracy). Specifically, GPT-4 frequently exhibited errors such as axis inversions, a phenomenon previously discussed in literature \cite{podo2024vi}. Despite this, both models demonstrated comparable accuracy,  indicating comparable efficacy in generating meaningful visualizations. 

\subsection{Qualitative evaluation}
\label{qualitative}

\begin{figure*}[!ht]
    \centering
    \includegraphics[width=\linewidth]{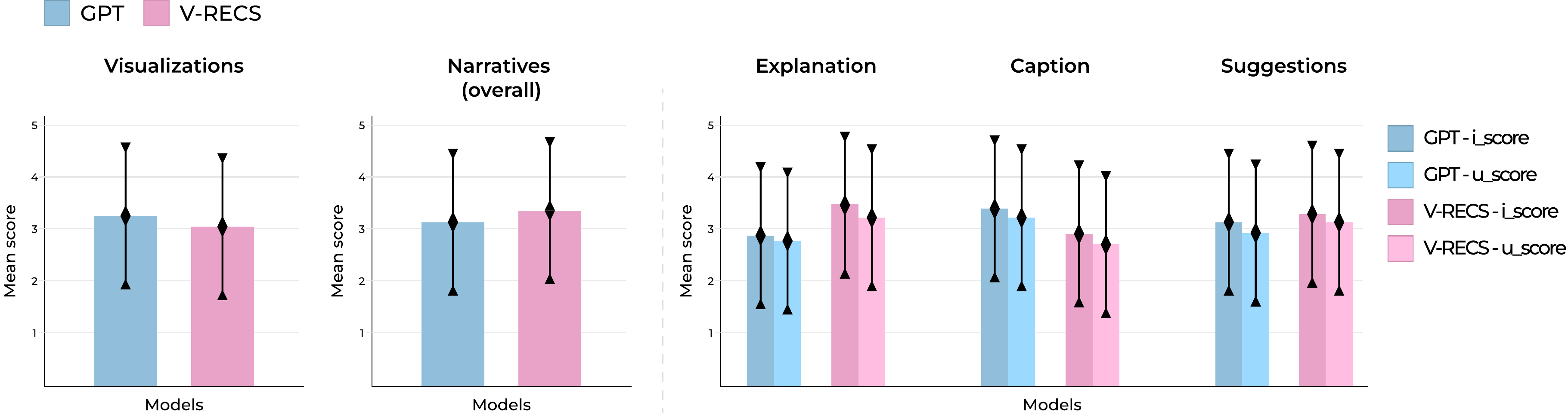}
    \caption{Average rank of the qualitative evaluation on a 1-5 scale. The first histogram to the left evaluates the visualization V, the second shows the overall usefulness of the visualization narrative  VN. The three subsequent histograms separately evaluate E, C, and S. For each type of narrative, participants have been asked to consider both the informativeness (i\_score)  and usefulness (u\_score). }
    \label{fig:userstudy-results}
\end{figure*}

\begin{figure}[!ht]
    \centering
    \includegraphics[width=0.85\linewidth]{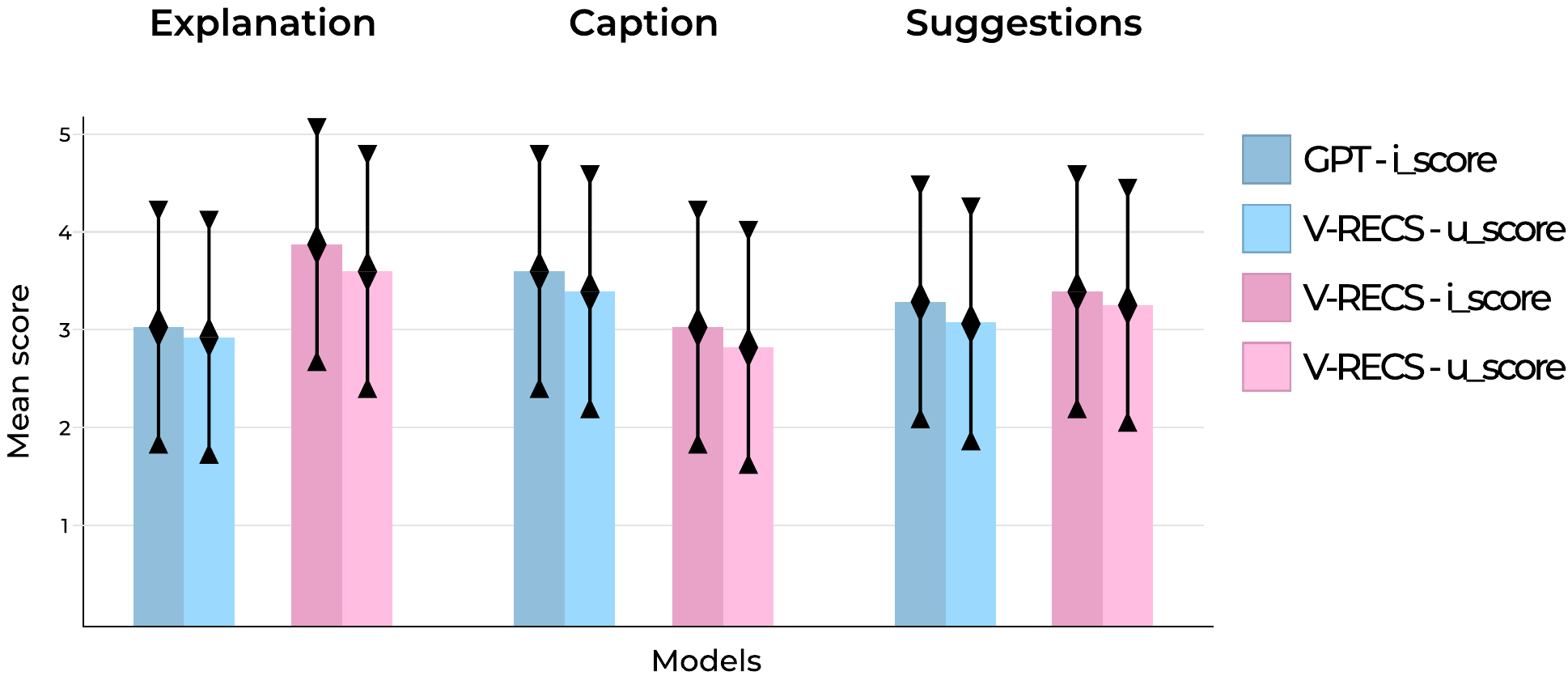}
    \caption{
    Narratives evaluation for the 16 non-expert participants. Results are confirmed even for this cohort, with a widening in performance in favor of V-RECS for the explanation narrative.}
    \label{fig:userstudy-non-expert}
\end{figure}

Qualitative evaluation involves the Informativeness and Effort layers. 
For the Informativeness layer, we have designed a qualitative experiment involving 50 randomly selected samples (not used for training) evaluated by 35 human annotators (27 males, 8 females, average age$= 29.6$ years, $std=9.8$). It followed an initial study with 15 participants where we mainly assessed the quality of the generated visualizations and the supporting narratives altogether. We maintained the original query difficulty distribution (as categorized in the source dataset), ensuring that our evaluation was representative of real-world scenarios. In addition to 50 examples of (D,Q) pairs extracted from NvBench, we also added ten slightly more natural sounding and complex sentences, to test the system ''into the wild". 
To facilitate the evaluation process, we developed a web application designed specifically for conducting a comparative study \footnote{\url{https://vrecseval.streamlit.app/}}. This application allowed us to present each volunteer with pairs of responses from the two models for the same query and dataset. The identity of the models' responses was hidden from participants, ensuring unbiased evaluations.
Each participant was tasked to evaluate the generated visualizations to assess their quality; for each of the associated narratives (E, C, and S), the participant was asked to express its informativeness and utility in supporting the interpretation of the visualization. We finally asked for an overall judgment on the utility of the narratives as a whole. All the responses used a Likert scale, ranging from 1 (Completely Meaningless) to 5 (Completely Meaningful). This scale provided a structured framework for capturing the nuanced differences in the quality and relevance of the responses.
Firstly, each participant was assigned a subset of 10 samples from the total pool of 60. This ensured that each sample received multiple evaluations from different participants, thereby mitigating individual biases.
To minimize the impact of any outliers or inconsistencies in the evaluations, we enforced a minimum threshold of three unique evaluations per sample. 

Figure \ref{fig:userstudy-results} shows the results of the evaluation. 
Looking at the first two charts on the left, we note how GPT4 and V-RECS present comparable performance for both visualization generation quality and overall narrative quality. Interestingly, V-RECS is slightly outperformed by GPT-4 for visualization quality while slightly outperforming GPT-4 in narrative quality.
The narratives analysis was further investigated individually, with results visible in the right part of figure \ref{fig:userstudy-results}.
Interestingly, V-RECS outperforms GPT-4 for explanation quality in both informativeness (VRECS: $\mu=3.54$, $std=1.40$; GPT-4: $\mu=2.91$, $std=1.31$) and utility (VRECS: $\mu=3.19$, $std=1.49$; GPT-4: $\mu=2.74$, $std=1.38$). This tells us that V-RECS is more capable of explaining the generation process to a user, helping in fostering more trust in the resulting visualization.
An inverted situation happens for the captioning narratives, with GPT-4 proving more capable than V-RECS at summarizing the content of a generated visualization. We believe that the generally richer captions generated by GPT-4 felt more informative and useful to the users. Due to a caption being a summary description of what a visualization shows, this result may be partially explained by the better capabilities of larger models like GPT-4 to produce effective summaries.


Finally, looking at the suggestions for further analysis narrative, V-RECS presents a slight edge on both perceived informativeness (VRECS: $\mu=3.37$, $std=1.18$; GPT-4: $\mu=3.16$, $std=1.23$) and utility (VRECS: $\mu=3.15$, $std=1.25$; GPT-4: $\mu=2.89$, $std=1.29$). Overall, the behavior on the S narrative presents comparable results between the two models.
We further analyzed the narratives by looking at the sub-cohort of nonexpert users (15 out of 35 declaring a degree of expertise ranging from none:1 to average:3, where average means the knowledge of basic visualization techniques and their usage in current workflow analyses).
Figure~\ref{fig:userstudy-non-expert} shows the results. Performance of V-RECS and GPT-4 remains confirmed for narratives C and S, with just a slight increase for both models on all the scores.
More interestingly, the differences in the Explanation quality are increased for both informativeness (VRECS: $\mu=3.78$, $std=1.09$; GPT-4: $\mu=2.98$, $std=1.11$) and utility (VRECS: $\mu=3.32$, $std=1.23$; GPT-4: $\mu=2.69$, $std=1.14$), still in favor of V-RECS. This result supports the capability of V-RECS to provide help to non-expert users in visualization for data analysis tasks.
Finally, considering the potential effect of the query difficulty, the results of this study confirmed that the quality of answers worsens with the complexity of natural language queries for both models, with V-RECS performing slightly better except for the extra hard sentences where results are comparable between the models. For the sake of space, the evaluation by complexity levels is reported in the supplemental material, along with other data.

The overall results of the manual evaluation, reported in aggregated form, have been added to the Narrative importance level of the informativess layer, as depicted in Figure \ref{fig:evallm}.

To conclude the qualitative analysis, we consider the LLM effort and the efficiency aspects of our evaluation. 
Efficiency, depicted for all dimensions in Figure \ref{fig:evallm} (right), entails comparing the accuracy achieved by each model at each level, while also considering the model complexity. Notably, our findings suggest that results comparable to GPT-4 can be achieved with a much smaller and more manageable model, which is a relevant advantage in practical deployment scenarios.

On the other hand, V-RECS' implementation effort (see the LLM effort level in Figure \ref{fig:evallm}) demands significantly more work compared to GPT. While GPT relies primarily on prompting techniques, V-RECS requires extensive fine-tuning, increasing the effort needed to achieve results comparable to GPT.


\subsection{Error Analysis}
\label{subsec:error analysis}

\begin{figure*}[!t]
\centering
\subfloat[]{\includegraphics[width=2.5in]{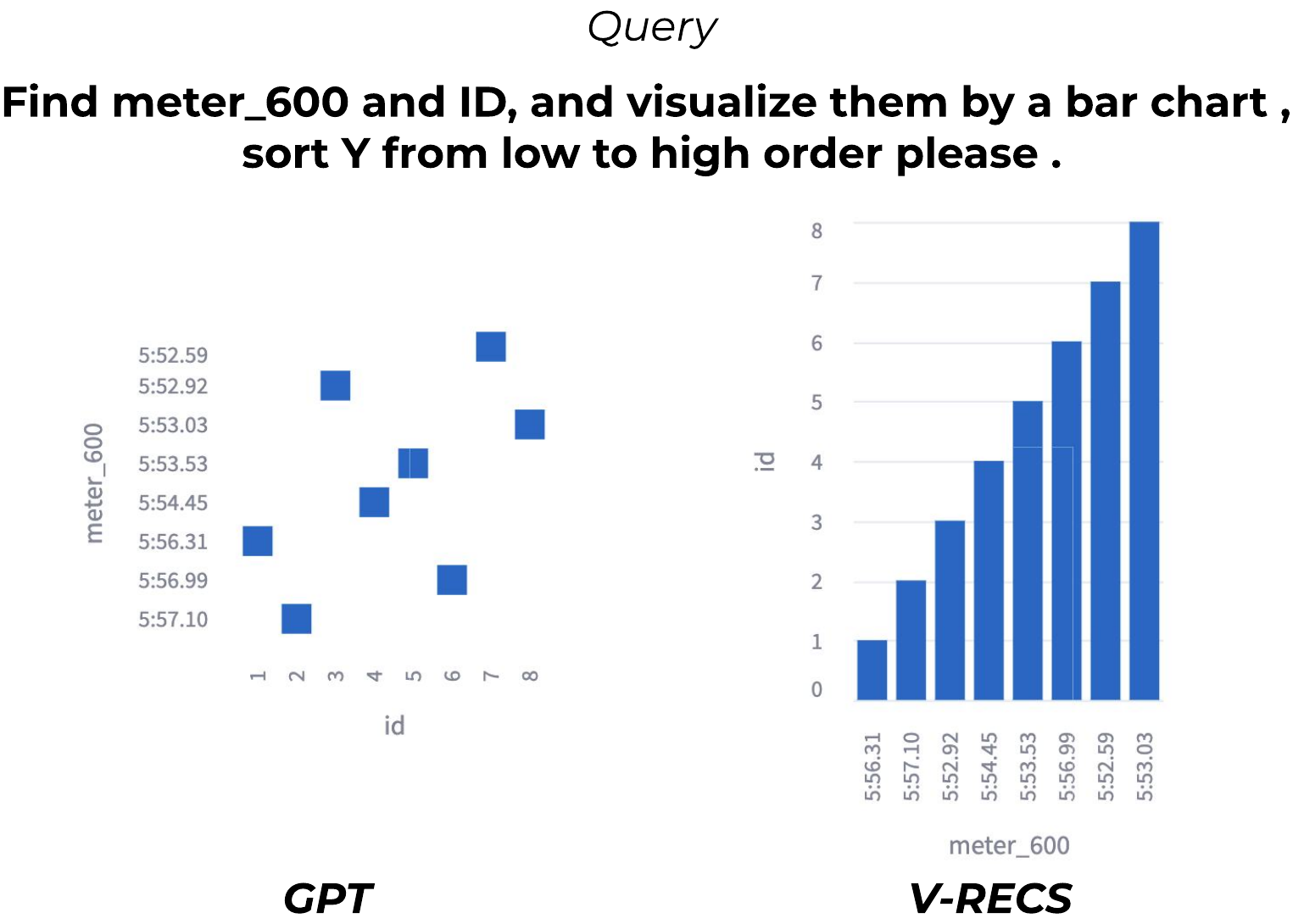}%
\label{fig:error1}}
\hfil
\subfloat[]{\includegraphics[width=2.5in]{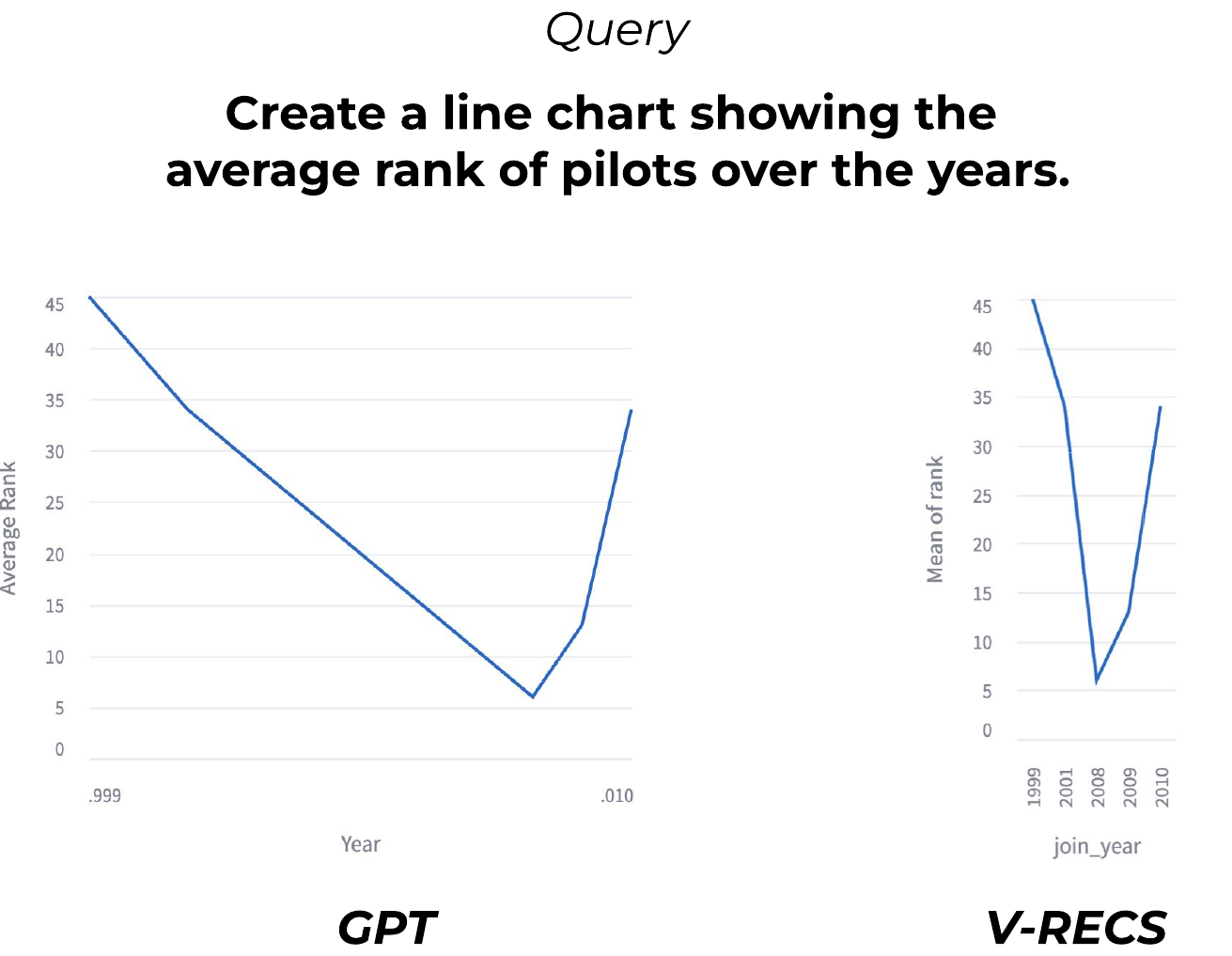}%
\label{fig:error2}}
\caption{Examples of errors in both systems. (a) Example of hallucination (GPT hallucinates). (b) Example of incorrect scaling (GPT) and spacing(V-RECS)}
\label{fig:errors}
\end{figure*}

Based on human evaluation and subsequent analysis by the authors, we detected common issues in V-RECS and GPT-4, comparing their outputs. 
Visualization errors in both systems can be categorized as follows:

\begin{enumerate} 
	\item \textbf{Incorrect Scaling:} Both systems sometimes struggle to scale the axes correctly, resulting in unusual and confusing data presentations. For example, in Figure \ref{fig:error2}, GPT-4 erroneously encoded years as numeric quantities, resulting in a misleading x-axis. 
	\item \textbf{Inverted Axes:} Sometimes, the models map columns to the incorrect axis, resulting in inverted axes. For example, V-RECS in Figure \ref{fig:error1} shows such an error. 
	\item \textbf{Non-Optimal Zoom/Spacing:} Both systems fail to optimize the visualization zoom, leading to charts where data is overlapped in a small section while a large portion remains empty and white. 
	\item \textbf{Hallucinations:} GPT-4 sometimes generates misleading visualizations. 
 As shown in Figure \ref{fig:error1}, GPT-4 created a new bar chart version with no meaningful visual representation, violating standard visualization principles. 
	\item \textbf{Missing Data:} Both models occasionally generate correct Vega-Lite specifications but fail to map the data, resulting in an empty grid and visualization. 
 \item  \textbf{Input errors} For a limited number of (Q,D) pairs, we noted that either Q is hardly interpretable even by a human, or missing/wrong data formats are present in D. This is due to problems in the NvBench dataset, already highlighted in Section \ref{subsec:implementation}. Surprisingly, although these problems lead to wrong visualizations, both systems are often still able to interpret the user's information need with substantially correct narratives.
\end{enumerate} 

All discussed errors are common to both models, except hallucination, which we detected only in GPT-4. 
We postulate that V-RECS is less prone to hallucinations due to its more straightforward grammar structure for generating the visualizations. VegaLite allows users to customize visualizations extensively, which can sometimes result in misleading visual representations. Conversely, V-RECS is built upon VegaZero, a streamlined version of VegaLite. VegaZero reduced the complexity by reducing the customization options, ensuring that the focus remains on the core properties of the visualization, thereby mitigating the risk of generating erroneous visual content.

The above-listed issues highlight existing challenges in ensuring the reliability of automated visualization systems and underscore the need for further improvements in their design and implementation.


\begin{table}[ht]
\caption{Comparison of GPT-4 and V-RECS visualization narratives (VN) for the query:\textbf{\textit{"What is the average rank of pilots in different positions?"}}. \textcolor{violet}{V-RECS narratives are colored in violet}, \textcolor{blue}{and GPT-4 narratives are colored in blue.}}
\label{tab:suggestions}
\fontsize{7pt}{9pt}\selectfont
\begin{tabularx}{\linewidth}{lX}
\toprule
VN & Responses \\ 
\midrule
E &
\textcolor{violet}{The 'position' and 'rank' columns are selected because they directly answer the user's question. The 'position' column provides the different categories (pilots) and the 'rank' column provides the numeric data to be averaged. The 'mean' aggregate function is used to calculate the average rank for each position. The 'group' transform is applied to group the data by position, which allows us to calculate the average rank for each position. Other columns like pilot id, pilot name, age, nationality, join year, and team are not selected as they do not contribute to answering the user's question.}
\textcolor{blue}{The 'rank' and 'position' features are the best among all the others from the dataset based on the user instruction because the user wants to know the average rank of pilots in different positions The 'rank' feature provides the rank of the pilots and the 'position' feature provides the different positions of the pilots} \\
\midrule
C & 
\textcolor{violet}{The visualization is a bar chart where the x-axis represents the 'Position' and the y-axis represents the 'Rank'. The height of each bar corresponds to the mean rank for each position. The bars are grouped by 'Position'.}
\textcolor{blue}{The visualization is a bar chart that shows the average rank of pilots in different positions. The x-axis represents the different positions and the y-axis represents the average rank. Each bar corresponds to a position and the height of the bar represents the average rank of pilots in that position} \\
\midrule

S & 
\textcolor{violet}{
1) Which position has the highest average rank?
2)Which position has the lowest average rank?
3) Are there any positions where the average rank is significantly different from others?}

\textcolor{blue}{
1) What is the average age of pilots in different positions?
2)What is the average join year of pilots in different positions?
3)What is the distribution of ranks in different positions?
}\\
\bottomrule

\end{tabularx}
\end{table}

As far as the E,C,S narratives are concerned,  we noticed that the explanations and captions produced by the two systems are slightly different but equally satisfactory, with V-RECS producing richer explanations and GPT better captions, as also highlighted by the human evaluation in Section \ref{qualitative}. Specifically,  V-RECS provides a more comprehensive explanation of the reasoning steps as shown in the example of Table \ref{tab:suggestions}. This result mitigates some of the concerns reported in the user study by Kim et al.~\cite{kim2024good} in terms of lack of depth and critical thinking capabilities while explaining a visualization design. On the other hand, despite judges have considered V-RECS suggestions slightly more useful to explore the recommended visualization, we noted that GPT-4 suggestions go beyond the user's query, presenting greater diversity and serendipity, as shown in the example of Table \ref{tab:suggestions}. We believe this feature (adding suggestions for additional exploration of the dataset rather than solely on the presented visualization)  could be implemented in our future research by improving the fine-tuning prompt strategy for the T3 task. 
%

\section{Conclusions and Future work} 
\label{sec:discussion}

This paper has proposed V-RECS, the first LLM-based visual recommender enhanced with narratives, and has demonstrated its capabilities.
The proposed model mitigates, among others, two relevant challenges of state-of-the-art LLM4Vis systems, discussed in Section \ref{sec:introduction}: lack of verifiability and controllability. Concerning the first, V-RECS allows both expert and non-expert users to evaluate the quality and significance of a proposed visualization through a rich narrative composed of an explanation and a caption. In addition, it also suggests additional queries for further data exploration.
Concerning the second challenge, V-RECS is based on fine-tuning a ``small'', open-source model, Llama-2-7B, while achieving comparable performance to a much larger, proprietary model, GPT-4, as we demonstrated in Section \ref{sec:evaluation}.

At the same time,  we list hereafter some limitations that V-RECS presents to encourage further research on the topic and opportunities that the availability of V-RECS enables for visualization and visual analytics researchers and practitioners.

\subsection{Limitations and challenges}
\noindent{\textbf{Visualization Literacy Coverage}}: The fine-tuning of V-RECS and quantitative experimentation are at this moment based on the NCnet dataset, a derivation from the well-known and state-of-the-art NVBench dataset.
As discussed in Section \ref{subsec:implementation}, this choice was made as this is the only dataset capable of supporting the generation and evaluation of the visualization along with its related narratives E,C,S.
Although additional datasets have been  made available, 
both in terms of chart collection methods discussed by Chen and Liu in their recent state-of-the-art report on corpora for automating chart generation~\cite{Chen2023}, or the recently available VisText dataset for benchmarking captioning quality~\cite{2023-vistext} these datasets are not readily usable for our purposes, as discussed in Section \ref{subsec:implementation}. 
We foresee the need for an effort
to create an integrated resource capable of evaluating all the features produced by V-RECS in a combined way.\\
\noindent{\textbf{User Evaluation}}: In its current form, the qualitative evaluation of V-RECS has been conducted through a comparative experiment with 35 annotators. We foresee the need to expand this study to a broader set of participants to explore the subtle differences that could exist in the presentation and significance layers of the EvaLLM evaluation stack and better evaluate the difference in our method between the proprietary GPT model and the open-source fine-tuned model.
Given the richness of this activity, we foresee it as a dedicated effort to be conducted in future work.\\
\noindent{\textbf{Form of Narrative}}: In its current form, the explanation (E) provided to illustrate the generation process is reported only in textual form.
As much as it is the standard for LLMs, and V-RECS aligns with it, we see as a limitation the lack of visual support to this explanation or lack of coordination between sub-parts of the explanation and visual elements of the generated chart. In this sense, the exploration of visual methods to support the explanation, as the ones reported by La Rosa et al.~\cite{LaRosa2023} or recently by Spinner et al.~\cite{Spinner2024} for LLMs, should be the area for further enhancement of this approach. In addition to that, we note that the answer to a specific query (both the chart and the narrative) is currently unique, while in principle several alternatives could be presented to the user, avoiding the illusion that the output is exclusive, and potentially revealing
inconsistencies. The latter is a relevant potential area for further improvement.

\subsection{Opportunities}
Along with limitations, we also report opportunities that V-RECS enables for the visualization community:\\
\noindent{\textbf{Support to democratization of powerful LLMs}} The methodology followed for the creation of V-RECS can be reapplied to fine-tune a new model for different tasks in the visualization field, such as visualization modifications or generation of a bitmap version of a visualization, exploiting a large foundation model as the Teacher model (e.g., ChatGPT for text-to-text, or DALL-E for text -to-image) at a relatively low-cost. In this way, the capabilities of these larger models can be inherited (evaluating the degree of inheritance correctly) and used, analyzed, or modified by researchers and practitioners, fostering their spread and better comprehension.\\
\noindent{\textbf{Support to visualization and visual analytics pipelines}}
V-RECS, the resulting model from the application of the proposed methodology, can be exploited for usage in visualization pipelines and visual analytics pipelines as a building block of the proposed system: looking at the former opportunity, tasks described by Schlesinger et al.~\cite{Schetinger2024} such as Chart recommendation and Rapid prototyping are already supported by V-RECS as is, while others such as the generation of Moodboards represent an easy extension. Interestingly, by construction V-RECS can also support tasks like visualizing the effect of training data on the outcomes of the model providing overall more controllability in the visualization recommendation process.
Focusing on the latter, a very recent work by Zhao et al.~\cite{Zhao2024}, LEVA, systematized the usage of LLM for Visual Analytics through the definition of three main stages of intervention: onboarding, exploration, and summarization. In this context, V-RECS fits the summarization stage for generating visual summarization and explanations, eventually proposing directions to proceed with the analysis through its suggestions (S). It could even be adapted to the other two stages through the application of the methodology for generating a path of analysis (check the previous opportunity) by researchers or practitioners.\\
\noindent{\textbf{Support for training activities}}
Still referring to the work by Zhao et al.~\cite{Zhao2024}, we foresee the opportunity to exploit V-RECS for training activities (referred to as the onboarding stage for Visual Analytics systems) on how to develop a visualization for a specific analysis task, through the support of explanations, or a workflow, exploiting a chain in which at each recommendation stage a new suggestion is selected and given as input for the next recommendation step.
This activity would be in line with the recent trend of using LLM for education, like the experiment by Chen et al.~\cite{Chen2023b} testing ChatGPT-3.5 and 4 to complete assignments of a visualization course (CS171) and the preliminary work proposed by Joshi et al.~\cite{joshievaluating} on leveraging ChatGPT-3.5 and Bard 2.0 for teaching a visualization technique.




\section*{Acknowledgments}
The authors wish to thank the participants who contributed to the manual assessment of V-RECS.
This work was in part supported by a grant from Lazio Region, FESR Lazio 2021-2027, project @HOME (\# F89J23001050007  CUP B83C23006240002).
This research has also been supported by the AWS Cloud Credit for Research Program, and the OpenAI’s Researcher Access Program.

\bibliographystyle{IEEEtran}
\bibliography{bibliography}

\section{Biography Section}

\begin{IEEEbiography}[{\includegraphics[width=1in,height=1.25in,clip,keepaspectratio]{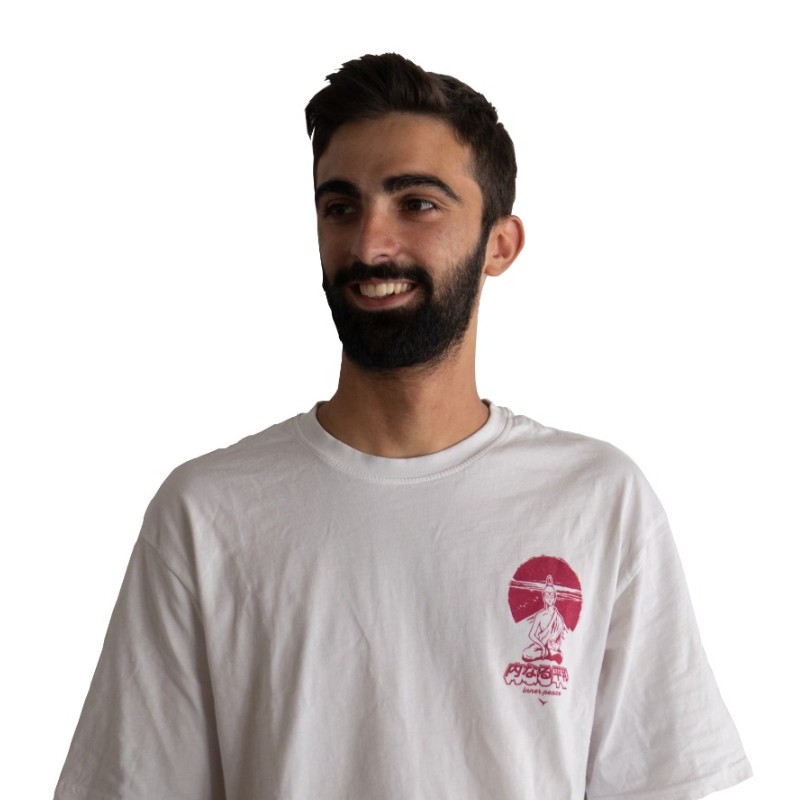}}]{Luca Podo} is a PhD Student in ML4VIS in the Sapienza University of Rome. He completed his MSc in Computer Science from Sapienza University of Rome in 2021. He founded the DeepViz Lab within the Computer Science Department at Sapienza, a group of young students and researchers in prototyping intelligent visual analytics systems. His research interests span Machine Learning for Visualizations, Visual Analytics, and Large Language Models.
\end{IEEEbiography}

\begin{IEEEbiography}[{\includegraphics[width=1in,height=1.25in,clip,keepaspectratio]{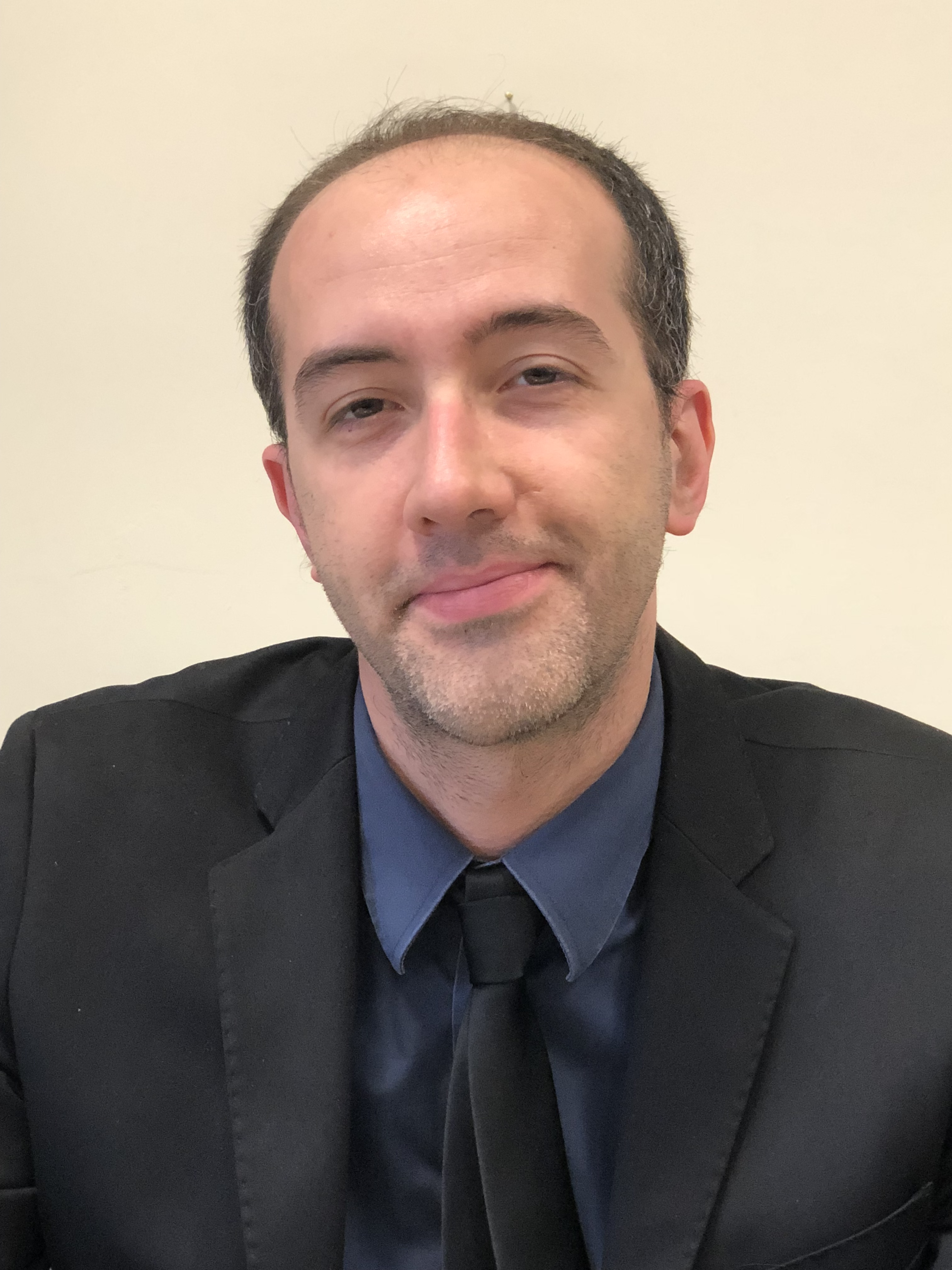}}]{Marco Angelini}
is an associate professor in computer science with Link University Rome.
He is a member and coordinates research projects of the A.W.A.RE group at Sapienza University of Rome. His main research interests include visual analytics, progressive visual analytics, and human-centered AI. More about him can be found at: \url{https://sites.google.com/dis.uniroma1.it/angelini}
\end{IEEEbiography}

\begin{IEEEbiography}[{\includegraphics[width=1in,height=1.25in,clip,keepaspectratio]{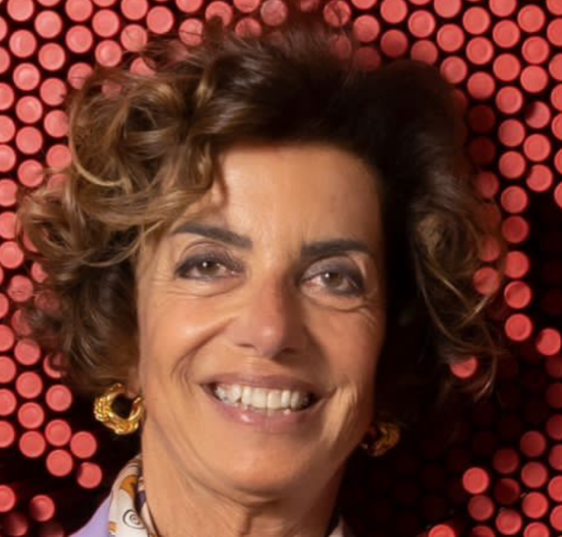}}]{Paola Velardi} is a full professor with the Department of Computer Science at Sapienza University of Rome. She has been working on Artificial Intelligence since 1983 as a visiting researcher at Stanford University. Her main research interests are text processing, machine learning, and knowledge bases.  Applications of interest include social networks, e-health,  e-learning, bioinformatics, recommender systems, cultural heritage, and machine learning for visual analytics. She published over 200 papers in all major scientific journals and venues. She has also designed and coordinated many projects to bridge the gender gap in ICT.
\end{IEEEbiography}

\vfill

\end{document}